\documentclass[pdflatex,sn-mathphys-num]{sn-jnl}
\usepackage{float}
\usepackage{siunitx}
\usepackage{verbatim}
\usepackage{graphicx}
\usepackage{amsmath}
\usepackage[usenames]{color}
\usepackage{amssymb}
\usepackage{hyperref}
\newcommand{\be}{\begin{equation}}
\newcommand{\ee}{\end{equation}}
\newcommand{\bea}{\begin{aligned}}
\newcommand{\eea}{\end{aligned}}
\newcommand{\pr}{\partial}

\newcommand{\bse}{\begin{subequations}}
\newcommand{\ese}{\end{subequations}}

\newcommand{\rb}{\bf r}
\newcommand{\xb}{\bf x}

\newcommand{\km}{\mathrm{k}}

\renewcommand{\v}[1]{\ensuremath{\mathbf{#1}}} 
\newcommand{\bmm}{\begin{multline}}
\newcommand{\emm}{\end{multline}}

\newcommand{\mi}{\mathrm{i}}
\begin{document}
\title{Superradiant scattering of electromagnetic fields from ringing black holes}
\author*[1]{\fnm{Rajesh} \sur{Karmakar}}\email{rajesh018@iitg.ac.in}

\author[2]{\fnm{Debaprasad} \sur{Maity}}\email{debu@iitg.ac.in}
\equalcont{These authors contributed equally to this work.}


\affil[1,2]{\orgdiv{Department of Physics}, \orgname{Indian Institute of Technology Guwahati}, \orgaddress{
\city{Guwahati}, \postcode{781039}, \state{Assam}, \country{India}}}



\abstract{
Detection of gravitational waves (GWs) paves the beginning of a new era of gravitational wave astronomy. Black holes (BHs) in their ringdown phase (ringing BHs) provide the cleanest signal of emitted GWs that imprint the fundamental nature of BHs under low energy perturbation. Apart from GWs, any complementary signature of ringing BHs can be of paramount importance. Motivated by this, we analyzed the scattering of electromagnetic waves in such a background and demonstrated that the absorption cross section of a ringing Schwarzschild BH can be superradiant. It appears that superradiance in the ringdown phase results from the stimulated growth of the matter field, driven by the oscillating gravitational background. Therefore, the amplification is inherently transient, with a characteristic time scale equal to the GW oscillation time scale. We have further analysed the frequency ranges of such amplified transient signals for a wide range of BH masses. In this context, we propose the premise that the primordial black hole (PBH) may undergo a merging phase and exhibit superradiance. We point out that the existing ground-based Low Frequency Array (LOFAR), radio telescopes could be able to detect such transient signals from PBHs, with mass range $M\sim 10^{-1} - 10^{-2} M_{\odot}$, going through the ringdown phase. Our present result, therefore, opens up an intriguing possibility of observing the PBH-PBH merging phenomena through electromagnetic waves.}

\maketitle

\newpage
\section{Introduction}\label{intro} 
The past few years have been very exciting in the field of gravity after the detection of GWs by LIGO-Virgo-KAGRA (LVK) \cite{LIGOScientific:2016aoc, LIGOScientific:2016emj, LIGOScientific:2016vbw, LIGOScientific:2016vlm, LIGOScientific:2016sjg, LIGOScientific:2017ycc, LIGOScientific:2017vwq, LIGOScientific:2020stg, LIGOScientific:2016jlg, LIGOScientific:2020aai, LIGOScientific:2020zkf, LIGOScientific:2017vox, LIGOScientific:2023fpk, KAGRA:2023pio, KAGRA:2022twx} which significantly enhanced our understanding of the nature of gravity and BHs in particular. The merging of binary BHs is a remarkable cosmic event, and the detection enables us for the first time to see those hidden phenomena through their gravitational wave emission. The whole merging process of two BHs usually consists of three distinct phases, inspiral, merger and ringdown (for binary BH observations see \cite{LIGOScientific:2016aoc, LIGOScientific:2016jlg}, for numerical simulations, one may look at \cite{Blackman:2017pcm, Centrella:2010mx, Boyle:2007ft, Ajith:2007kx}). In our present study, we particularly focus on analysing characteristic features of a BH in its ringdown phase (which we shall refer to as a ringing BH) when interacting with external fields. In the previous study \cite{Karmakar:2021ssg}, to the best of our knowledge, we for the first time investigated the scattering of an ultralight scalar field with a ringing Schwarzschild BH and demonstrated that the field might experience superradiant scattering, particularly in the infrared regime. However, the detection of such transient superradiant signals in the scalar sector is experimentally challenging. In this paper, we investigate instead the scattering of electromagnetic (EM) fields, which is experimentally more relevant. Apart from direct gravitational waves, any complementary signature such as time-dependent superradiance scattering of EM waves from ringing BH could be of paramount importance in light of the recent spate of research activity on the aspects of black holes, and gravitational waves in particular. Observation of gamma-ray bursts from binary neutron star merger, GRB 170817A, by Fermi-GBM \cite{Goldstein:2017mmi} played such a complementary role in association with the detection of GWs by LVK \cite{LIGOScientific:2017vwq} to establish the source of these radiations on a firm footing.

EM waves interacting with a gravitational wave have been explored in the literature\cite{Patel:2021cat, Clarkson:2004, Benone:2019all}. Phenomena of superradiance, i.e., amplification of scattered waves, were first initiated by Zel'Deovich\cite{Zel’Dovich1971, Zel’Dovich1972} in the context of a rotating body. Later on such analyses have been carried forward by Misner\cite{Misner1972}, Bekenstein\cite{Bekenstein:1973mi}, and Starobinsky \cite{Starobinskii:1973hgd, Starobinskil:1974nkd} to study the energy extraction process from the black hole and its stability against the external perturbations, which have been the subject of investigation for quite some time\cite{Crispino:2007qw, Crispino:2008zz, Leite:2018mon, Leite:2017zyb}, particularly for rotating BHs.
The possibility of moving BHs \cite{Cardoso:2019dte}, causing superradiant amplification, has also been explored in the literature. On the observational front, the dynamics of fundamental fields in the time-dependent background have been the subject of investigation in recent times\cite{Chen:2023vkq}, where the superradiant evolution leads to a significant enhancement in the emitted flux. The dynamical character of the superradiance very often gives rise to interesting effects in the BH shadow\cite{Roy:2019esk, Roy:2021uye} and polarization \cite{ Chen:2019fsq, Yuan:2020xui}. In this paper, we initiate a novel study of electromagnetic wave scattering due to the ringing BH background. 

The strategy to compute the absorption cross section in a time-dependent background is not well established. This task is more challenging for a ringing BH given that the spacetime structure is not asymptotically flat. In our previous work \cite{Karmakar:2021ssg}, we have developed a methodology to evaluate the absorption cross section of such BH spacetime for a scalar field by introducing a hypothetical interaction surface, where the matter field interacts with the GW wave background. The absorption cross section naturally got parametric dependent characteristics based on the position of this surface. In the present analysis, we extended this study to an electromagnetic field and demonstrated a strategy to circumvent the issues with such dependence on the said parameter by proposing an average value of the absorption cross section.

 As an application of our model, we have explored the potential scenario for merger events involving primordial black holes (PBHs), taking advantage of the wide range of possible PBH masses \cite{Raidal:2017mfl}. Recent GW observations by LVK \cite{LIGOScientific:2016aoc, LIGOScientific:2016emj, LIGOScientific:2016vbw, LIGOScientific:2016vlm, LIGOScientific:2016sjg, LIGOScientific:2017ycc, LIGOScientific:2017vwq, LIGOScientific:2020stg, LIGOScientific:2016jlg, LIGOScientific:2020aai, LIGOScientific:2020zkf, LIGOScientific:2017vox, LIGOScientific:2023fpk, KAGRA:2023pio, KAGRA:2022twx} have revealed that BHs possess an extended mass distribution. Therefore, PBHs could serve as plausible sources of GWs detected by LVK, as they are capable of undergoing the inspiral-merger-ringdown process (see \cite{Sasaki:2018dmp} for current constraints on PBHs in light of GW observations). This has driven us to investigate the detectability of the PBH-PBH mergers through other fundamental fields apart from GWs, as such observation would provide complementary evidence for the merger process.

The rest of the paper is organized in the following manner. First, we briefly describe the ringing BH spacetime in Sec.\ref{ringing_spacetime} (details can be found in \cite{Karmakar:2021ssg, Zerilli:1971wd}). In Sec.\ref{eom_gauge}, we derive the governing equation to study the dynamics of the EM field in this background and subsequently present how the gravitational quasinormal mode frequencies manifest in the dynamical equations of the gauge field in Sec.\ref{Coupling_frequency}. In Sec.\ref{bc} we establish the framework to calculate the absorption cross section by introducing a hypothetical interaction surface. This surface represents the position where incoming radiation interacts with the oscillating gravitational wave background. By fixing suitable boundary conditions of the EM field, we then discuss, in Sec.\ref{def.ACS}, the procedure to evaluate the absorption cross section with approximate normalization suitable for the oscillating BH. Next, in Sec.\ref{num.results}, numerical results of the absorption cross section have been presented for individual interaction surfaces. In Sec.\ref{meanACS}, we have defined the mean absorption cross section by averaging over the position of the interaction surface and argued how the energy extraction takes place. Finally, we discuss the possible observational scenarios and conclude with future directions. 

We will use the metric signature as $(-,+,+,+)$ and follow the natural units, $\hbar=c=G=1$, throughout the discussion.
\section{Ringing Black Hole Background}\label{ringing_spacetime}
The final stage of a binary black hole merger, known as the ringdown phase, can be modelled theoretically by applying perturbation theory on a static BH spacetime. For our present analysis, the static part of the ringing BH is assumed to be Schwarzschild. In the perturbative limit, the ringing BH can be mathematically expressed in the following manner, $g_{\mu\nu}= {g^0_{\mu\nu}+h_{\mu\nu}}$, where $g_{\mu\nu}^0$ is the standard Schwarzschild metric. $h_{\mu\nu}$ is the gravitational perturbation, with $|h_{\mu\nu}|<< g^0_{\mu\nu}$. We further consider the quadrupole oscillation with orbital angular mode $l_0=2$, azimuthal angular mode $m_0=0$ (note that quasinormal mode frequency does not depend on $m_0$ \cite{Regge:1957td}), of the fluctuation part. This oscillating (ringing) part ( see Appendix A of \cite{Karmakar:2021ssg} for details) is  expressed in the radiation gauge \cite{Zerilli:1971wd}, which captures the correct asymptotic behaviour of the GW flux, as,
\be\label{ringingmetric}
h_{\mu\nu}=\frac{1}{2}e^{-\mi\omega{t}}\begin{pmatrix}
	Hf(r)Y_2^0 & H_1Y_2^0 & 0 & 0\\
	H_1Y_2^0 & Hf(r)^{-1}Y_2^0 & h^{(e)}_{1}\pr_\theta{Y_2^0} & h^{(o)}_1s_\theta\pr_\theta{Y_2^0}\\
	0 & h^{(e)}_{1}\pr_\theta{Y_2^0} & r^2\mathcal{T}_2^0 & \frac{1}{2}h_2\mathcal{I}_2^0\\
	0 & h^{(o)}_1s_\theta\pr_\theta{Y_2^0} & \frac{1}{2}h_2\mathcal{I}_2^0 & r^2s^2_\theta\tilde{\mathcal{T}}_2^0
\end{pmatrix} + c.c. 
\ee
Where, symbols are, $\mathcal{I}_2^0=(c_\theta\pr_\theta{Y_2^0}-s_\theta\pr^2_\theta{Y_2^0})$, 
$\mathcal{T}_2^0=KY_2^0+G\pr^2_\theta{Y_2^0}$, and
$\tilde{\mathcal{T}}_2^0=K Y_2^0+\cot\theta{G} \pr_\theta{Y_2^0}$. $s_\theta=\sin\theta$, $c_\theta=\cos\theta$ and $Y_{l}^{m}$ is the spherical harmonics. Note that we have added the complex conjugate (c.c.) part to make the fluctuation, $h_{\mu\nu}$, real. The time-dependent part of the ringing fluctuation is expressed as $ e^{-i \omega t}$, with $\omega$ being the quasi-normal mode frequency. The perturbation variables are divided into parity odd ($h^{(o)}_1,h_2$) and parity even $(h^{(e)}_1,H, H_1, K,G)$, radial coordinate $(r)$ dependent, functions. 
Einstein's equation governing the ringing perturbation variables boils down to the well-known Regge-Wheeler equations \cite{Regge:1957td, Edelstein:1970sk, Zerilli:1970se}
\be\label{regee-wheeler.eq}
\frac{d^2 \tilde{Z}_i}{dr_*^2}+(\omega^2-V_{i}){\tilde Z}_i=0 ,
\ee
where, $i\equiv ({\cal{E}},{\cal O})$ are associated with ${\cal{E}}$ven and ${\cal O}$dd perturbation. Whereas $r_* = r + 2 M \ln(r/2M -1)$ is the Tortoise coordinate. For quadrupole oscillation, the potentials assume the following form,
\begin{eqnarray}
V_{\cal{E}} &=& f(r) \frac{8(3r^3 + 3 Mr^2 )+18{M^2}(2r+M)}{r^3(2{r}+3M)^2}, \nonumber\\
V_{\cal{O}} &=& f(r)\Big(\frac{6}{r^2}-\frac{6M}{r^3}\Big).
\end{eqnarray}
Where, $f(r) =1-{2M}/{r}$ is the Schwarzschild metric function, and $M$ is the mass of the black hole. The functional dependence of odd parity variables $\tilde{Z}_{\mathcal{O}}(r)$ and even parity variables $\tilde{Z}_{\mathcal{E}}(r)$ are explicitly derived in \cite{Zerilli:1971wd, Zerilli:1970se}. The near horizon values of the ringing fields will be parameterized by $(|\tilde{Z}_{\mathcal{O}}(r\to{2M})|= {\cal O}_h,|\tilde{Z}_{\mathcal{E}}(r\to{2M})| = {\cal E}_h)$. By solving the Regge-Wheeler equations\eqref{regee-wheeler.eq} with the quasinormal mode boundary conditions \cite{Chandrasekhar:1975zza, Chandrasekhar:1975zzb}; ingoing near the event horizon and outgoing near spatial infinity, implementing shooting method \cite{Chandrasekhar:1975zza, Chandrasekhar:1975zzb, Molina:2010fb} in Mathematica, we have been able to find out the quasinormal mode eigenfunctions and construct the full solution of ringing Schwarzschild BH background \eqref{ringingmetric}. Next, we focus on solving the minimally coupled EM equation in such a background.  

\section{Minimally Coupled Gauge Field}\label{eom_gauge} 
The minimally coupled EM field satisfies the following equation of motion, 
\be\label{Eom}
\bea
&\pr_\mu(\sqrt{-g}g^{\mu\alpha}g^{\nu\beta}F_{\alpha\beta})=0 .
\eea
\ee
It is important to understand that one can always maintain the amplitude of the EM field, $[F_{\mu\nu}]< [{h^\mu}_\mu]$ with an overall multiplication by a small number as \eqref{Eom} remains invariant. This way we can make the energy momentum tensor of the EM field subleading as compared to the ringing background.To construct a linearized version of the equation of motion \eqref{Eom}, up to  $\mathcal{O}(h)$, we consider the determinant of the metric as, $\sqrt{-g}=\sqrt{-g_0}(1+{h^\mu}_{\mu}/2)$, with the trace of the metric taken to be ${h^\mu}_{\mu}=g^{\mu\nu}_0h_{\mu\nu}$. For the same purpose, we express the inverse metric as, $g^{\mu\nu}=g^{\mu\nu}_0-h^{\mu\nu}$. With this setup, the equation of motion takes the following form,
\be
\pr_\mu\left[\sqrt{-g_0}\left(1+\frac{1}{2}{h^\gamma}_{\gamma}\right)\big(g^{\mu\alpha}_0-h^{\mu\alpha}\big)\big(g^{\nu\beta}_0-h^{\nu\beta}\big)F_{\alpha\beta}\right]=0.
\ee
Up to linear order in fluctuation, $(\mathcal{O}(h))$, the above equation can then be expressed as, 
\be\label{linearEom}
\bea
&\pr_\mu(\sqrt{-g_0}g^{\mu\alpha}_0g^{\nu\beta}_0F_{\alpha\beta})+\pr_\mu\left[\sqrt{-g_0}\left(\frac{{h^\gamma}_\gamma}{2}g^{\mu\alpha}_0g^{\nu\beta}_0-g^{\mu\alpha}_0h^{\nu\beta}-h^{\mu\alpha}g^{\nu\beta}_0\right)F_{\alpha\beta}\right]=0.
\eea
\ee
Note that the expansion of the electromagnetic field tensor will be considered in the following analysis, once we express it in terms of the gauge field potential. Notice that the first part of the above equation describes the governing equation of the EM field for the static Schwarzschild BH, while the second part corresponds to that for the leading order ringing fluctuation.
Since the non-ringing part of the background is spherically symmetric, we decompose the EM field components as  \cite{Brito:2015oca}
\be\label{modedecom}
\bea
&A_t(t,\v r)=\sum_{lm} b^{lm}(t,r)Y_{lm}(\Omega) ,\\
&A_r(t,\v r)=\sum_{lm}d^{lm}(t,r)Y_{lm}(\Omega) ,\\
&A_{s}(t,{\rb})=\sum_{lm}\left[k_{lm}(t,r)\Psi^{lm}_{s}(\Omega)+a_{lm}(t,r)\Phi^{lm}_{s}(\Omega)\right],
\eea
\ee
where we have used the orthogonal vector spherical harmonic basis \cite{Barrera, Jackson:1998nia} for the azimuthal field components,
\be
\bea
&\Psi^{lm}_s=\pr_s{Y_{lm}},\\
&\Phi^{lm}_s=\epsilon_{ss'}\pr^{s'}{Y_{lm}} .
\eea
\ee
Here, and throughout the paper, $s,s'$ indices correspond to angular coordinates $(\theta,\phi)$. Levi-Civita symbols, $\epsilon_{\theta\theta}=\epsilon_{\phi\phi}=0,\epsilon_{\theta\phi}=-\epsilon_{\phi\theta}=\sin\theta$. 
In our analysis, we find it convenient to work with the following gauge invariant variables,
\be\label{invvar}
\bea
&\chi^{lm}_1=\frac{r^2}{l(l+1)}(\pr_t d^{lm}-\pr_r b^{lm})~~;~~\chi^{lm}_2=a^{lm} ,\\
&\chi^{lm}_3=d^{lm}-\pr_r k^{lm} ~~;~~\chi^{lm}_4=b^{lm}-\pr_t k^{lm},\\
\eea
\ee
which can be obtained from \eqref{modedecom}, and so constructed, they remain invariant under the gauge transformation of the EM field potential, $A_\mu$. Once the EM field tensor, $F_{\mu\nu}$, is expressed in terms of these gauge invariant variables, the next challenge in the governing dynamical equation \eqref{linearEom} arises from the spherical components, which appear as $Y_{lm}(\theta,\phi)Y_{20}(\theta,\phi)$, stemming from the quadrupole oscillations of the background. To decouple the spherical part we have made use of the following identity defined in terms of Wigner-3j symbols with the main definition given as,
\be
Y_{lm}(\theta,\phi)Y_{l'm'}(\theta,\phi)=\sum_{c\gamma}\Lambda^{(l',m')}_{lmc\gamma}Y_c^\gamma(\theta,\phi)
\ee
where, \[\Lambda^{(l',m')}_{lmc\gamma}=(-1)^{\gamma}\sqrt{\frac{(2l'+1)(2l+1)}{4\pi}}\sqrt{2c+1}\begin{pmatrix}
l & l' & c \\
m & m' & -\gamma\\
\end{pmatrix}\begin{pmatrix}
l & l' & c \\
0 & 0 & 0\\
\end{pmatrix}.\] For the non-zero value of the Wigner $3j$ coefficient 
$\begin{pmatrix}
l & l' & c \\
m & m' & -\gamma\\
\end{pmatrix}$,
all the $l$ values should satisfy the triangle law; the sum of any two $l$ values should be greater than or equal to the third one, and the sum of all the $m$ values should be zero. With this setup, we extract the governing equations for individual mode ($l,m$) from \eqref{chieqn} (after some tedious but straightforward mathematical steps), and express them as,
\be\label{chieqn}
\bea
&\mathcal{L}_0(t,r)\chi^{lm}_1+{\cal Q}^i_{lmc\gamma}(h)\chi^{c\gamma}_i+\bar{\cal Q}^i_{lmc\gamma}(h^*)\chi^{c\gamma}_i=0 ,\\
&\mathcal{L}_0(t,r)\chi^{lm}_2+{\cal R}^i_{lmc\gamma}(h)\chi^{c\gamma}_i+\bar{\cal R}^i_{lmc\gamma}(h^*)\chi^{c\gamma}_i=0,
\eea
\ee
with ${\cal L}_0(t,r)$ representing a Klein-Gordon operator for static Schwarzschild BH,
\be\label{sch.kg}
\mathcal{L}_0(t,r)=f(r)\pr_r(f(r)\pr_r) -\pr^2_t-f(r)\frac{l(l+1)}{r^2} ,
\ee
where $i\to(1,2,3,4)$ and all the repeated indices are assumed to be summed over. Whereas ${\cal Q}^i_{lmc\gamma},{\cal R}^i_{lmc\gamma}$ (for detailed expression see the appendix \ref{source}), are the differential operators depending on the first complex part of the fluctuation metric \eqref{ringingmetric}, while $\bar{\cal Q}^i_{lmc\gamma},\bar{\cal R}^i_{lmc\gamma}$ corresponds to the differential operators for the complex conjugate part (reason for $h^*$ in the parenthesis). Also, note that the new indices $c$ and $\gamma$ represent the orbital and azimuthal angular momentum modes, respectively, similar to $l$ and $m$. Nevertheless, the remaining two gauge-invariant variables $\chi^{lm}_{3}$ and $\chi^{lm}_{4}$ also lead to similar equations; however, it will not be required to find out the solution directly, as we will see in the later sections. 

The ringing background is constructed out of a quadrupole perturbation, and hence the spherical symmetry of the system under study is naturally lost. Consequently, different angular modes of the EM field are now coupled to each other as evident from the repeated indices, $i$, $c$ and $\gamma$ in \eqref{chieqn}. Therefore, to make the equation \eqref{chieqn} solvable we proceed by considering a perturbative expansion of the field variables, $\chi^{lm}_i$ as,
\be \label{expand}
\bea
\chi^{lm}_i(t,r)=\chi^{lm}_{i({0})}(t,r)+\chi^{lm}_{i({\rm p})}(t,r)+\mathcal{O}([h_{\mu\nu}]^2),
\eea
\ee
where the successive terms represent the gauge field in the order of $[h_{\mu\nu}]$. Field variables with subscript ``$(0)$" denote the solution for the static background, and the same with subscript ``$(\rm p)$" corresponds to the perturbed part. Substituting the above expansion in \eqref{chieqn} we obtain the linearized equations of motion, up to $\mathcal{O}([h_{\mu\nu}]^0)$,
\be\label{eom_expand1}
\bea
 &{\cal L}_0(t,r)\chi^{lm}_{1({0})} = 0, \\
 &{\cal L}_0(t,r)\chi^{lm}_{2({\rm 0})} = 0, \\.
\eea
\ee
and up to  $\mathcal{O}([h_{\mu\nu}])$,
\be\label{eom_expand2}
\bea
 &{\cal L}_0(t,r)\chi^{lm}_{1({\rm p})} +{\cal Q}^i_{lmc\gamma}(h)\chi^{c\gamma}_{i(0)}+\bar{\cal Q}^i_{lmc\gamma}(h^*)\chi^{c\gamma}_{i(0)}= 0, \\
 &{\cal L}_0(t,r)\chi^{lm}_{2({\rm p})} + {\cal R}^i_{lmc\gamma}(h)\chi^{c\gamma}_{i(0)}+\bar{\cal R}^i_{lmc\gamma}(h^*)\chi^{c\gamma}_{i(0)}= 0 .
\eea
\ee
Once again, we neglect the $\mathcal{O}([h_{\mu\nu}]^2)$ terms, given that the amplitude of the ringing fluctuation is very small as compared to the background. Nevertheless, by using the properties of the non-homogeneous differential equation, we now simplify the above equations, \eqref{eom_expand1} and \eqref{eom_expand2}, with the decomposition of the fields taken as, $\chi^{lm}_{1({\rm p})}= \chi^{lm}_{1,{\rm p}}+\bar\chi^{lm}_{1,{\rm p}}$ and $\chi^{lm}_{2({\rm p})}= \chi^{lm}_{2,{\rm p}}+\bar\chi^{lm}_{2, {\rm p}}$, such that,  
\be\label{part.eq1}
\bea
&{\cal L}_0(t,r)\chi^{lm}_{1,{\rm p}} + {\cal Q}^i_{lmc\gamma}(h) \chi^{c\gamma}_{i(0)}=0,\\
&{\cal L}_0(t,r)\bar{\chi}^{lm}_{1,{\rm p}}+\bar{\cal Q}^i_{lmc\gamma}(h^*)\chi^{c\gamma}_{i(0)}= 0,
\eea
\ee
and 
\be\label{part.eq2}
\bea
&{\cal L}_0(t,r)\chi^{lm}_{2, {\rm p}} + {\cal R}^i_{lmc\gamma}(h)\chi^{c\gamma}_{i(0)}=0,\\
&{\cal L}_0(t,r)\bar{\chi}^{lm}_{2,{\rm p}}+ \bar{ \cal R}^i_{lmc\gamma}(h^*)\chi^{c\gamma}_{i(0)}= 0 .
\eea
\ee
In the next section, we will present the above set of equations in the frequency domain, which is achievable in the perturbative approach discussed above.
\section{How the GW-QNMs manifest in the EM field}\label{Coupling_frequency} 
The previous set of equations, \eqref{part.eq1} and \eqref{part.eq2}, can be thought of as EM waves propagating in the static Schwarzschild ($g_{\mu\nu}^0$) background with oscillatory source term. Moreover, $\chi^{c\gamma}_{i(0)}(t,r)$ being the solution for static Schwarzschild BH, we can decompose it in Fourier space as,
\be\label{chi0time}
\chi^{c\gamma}_{i(0)}(t,r)=\int d\km~e^{-\mi {\rm k} t}\chi^{c\gamma}_{i(0)}({\rm k}, r) 
\ee
with ${\rm k}$ representing the corresponding frequency. Now, recall that the time dependence of the ringing fluctuation \eqref{ringingmetric} is dictated by $e^{-\mi\omega t}$, so that, one will be able to extract it out of ${\cal Q}_{lmc\gamma}$ and ${\cal R}_{lmc\gamma}$. While for the complex conjugate part, with $e^{\mi\omega t}$, the same is possible for $\bar{\cal Q}_{lmc\gamma}$ and $\bar{\cal R}_{lmc\gamma}$. The preceding two arguments help us to express \eqref{part.eq1} and \eqref{part.eq2} respectively as,
\small{
\be\label{kpart.eq1}
\bea
&{\cal L}_0(t,r)\chi^{lm}_{1,{\rm p}} + \int d\km~e^{-\mi {(\km+\omega)} t} {\cal Q}^i_{lmc\gamma}(\km,\omega)\chi^{c\gamma}_{i(0)}({\rm k},r)=0,\\
&{\cal L}_0(t,r)\bar{\chi}^{lm}_{1,{\rm p}}+\int d\km~e^{-\mi {(\km-\omega^*)} t}  \bar{\cal Q}^i_{lmc\gamma}(\km,\omega^*) \chi^{c\gamma}_{i(0)}({\rm k}, r)= 0,
\eea
\ee}
and 
\small{
\be\label{kpart.eq2}
\bea
&{\cal L}_0(t,r)\chi^{lm}_{2, {\rm p}} + \int d\km~e^{-\mi {(\km+\omega)} t}  {\cal R}^i_{lmc\gamma}(\km,\omega)\chi^{c\gamma}_{i(0)}({\rm k}, r)=0\\
&{\cal L}_0(t,r)\bar{\chi}^{lm}_{2,{\rm p}}+\int d\km~e^{-\mi {(\km-\omega^*)} t} \bar{ \cal R}^i_{lmc\gamma}(k,\omega^*)\chi^{c\gamma}_{i(0)}({\rm k}, r)= 0 .
\eea
\ee}
We only consider the particular solution for the $\chi^{lm}_{1,{\rm p}},\bar{\chi}^{lm}_{1,{\rm p}} $ and $\chi^{lm}_{2,{\rm p}},\bar{\chi}^{lm}_{2,{\rm p}}$, hence, it is plausible to consider the following decomposition in regard to the above sets of equations,
\be\label{chi1p.decomp}
\chi^{lm}_{1,{\rm p}}(t,r)=\int d\km~e^{-\mi {(\km+\omega)} t} \chi^{lm}_{1,{\rm p}}(\km, r)
\ee
and similar expressions for $\chi^{lm}_{2, {\rm p}}$ is understood. In the same fashion, we consider, 
\be\label{chi1pb.decomp}
\bar{\chi}^{lm}_{1,{\rm p}}(t,r)=\int d\km~e^{-\mi {(\km-\omega^*)} t}\bar{\chi}^{lm}_{1,{\rm p}}(\km , r)
\ee
and similar expressions for $\bar{\chi}^{lm}_{2, {\rm p}}$ is understood.

Substituting the above forms of the fields, we obtain from \eqref{kpart.eq1} and \eqref{kpart.eq2},
\be\label{tind.eq1}
\bea
&{\cal L}_0(\km,\omega,r)\chi^{lm}_{1,{\rm p}}(\km , r) + {\cal Q}^i_{lmc\gamma}(\km,\omega)\chi^{c\gamma}_{i(0)}(\km, r)=0,\\
&{\cal L}_0(\km,\omega^*,r)\bar{\chi}^{lm}_{1,{\rm p}}(\km, r)+\bar{\cal Q}^i_{lmc\gamma}(\km,\omega^*) \chi^{c\gamma}_{i(0)}(\km, r)= 0,
\eea
\ee
and 
\be\label{tind.eq2}
\bea
&{\cal L}_0(\km,\omega,r)\chi^{lm}_{2, {\rm p}}(\km , r) + {\cal R}^i_{lmc\gamma}(\km,\omega)\chi^{c \gamma}_{i(0)}(\km, r)=0\\
&{\cal L}_0(\km,\omega^*,r)\bar{\chi}^{lm}_{2,{\rm p}}(\km, r)+\bar{ \cal R}^i_{lmc\gamma}(\km,\omega^*)\chi^{c\gamma}_{i(0)}(\km, r)= 0.
\eea
\ee
Where, 
\be
\mathcal{L}_0(\km,\omega,r)=f(r)\pr_r(f(r)\pr_r) +(\km+\omega)^2-f(r)\frac{l(l+1)}{r^2} ,
\ee
and
\be
\mathcal{L}_0(\km,\omega^*,r)=f(r)\pr_r(f(r)\pr_r) +(\km-\omega^*)^2-f(r)\frac{l(l+1)}{r^2} ,
\ee
can be derived by acting $\mathcal{L}_0(t,r)$ \eqref{sch.kg} on \eqref{chi1p.decomp} and \eqref{chi1pb.decomp}.

Finally, we are left with \eqref{tind.eq1} and \eqref{tind.eq2}, which describe a nonhomogeneous differential equation with spatial variable, $r$. To solve these equations,  we are mainly focusing on the source contribution, i.e., the second term in each equation of \eqref{tind.eq1} and \eqref{tind.eq2}. Following the property of a non-homogenous differential equation, fixing the initial condition for $\chi^{lm}_{i(0)}$ would be sufficient, as the source term depends (along with the ringing metric components) on the zeroth-order solution of the EM field. To find out the zeroth order solution of all the gauge invariant variables, $\chi^{lm}_{i(0)}$ we solve $\chi^{lm}_{1(0)}(\km, r)$ and $\chi^{lm}_{2(0)}(\km, r)$ first by setting the ingoing boundary condition near the horizon of the static BH, as, $\chi^{lm}_{1(0)}(\km, r)|_{r\to 2M}=\zeta^{lm}_1 f(r)^{-2\mi M \km}$ and $\chi^{lm}_{2(0)}(\km, r)|_{r\to 2M} =\zeta^{lm}_2 f(r)^{-2\mi M \km}$, with arbitrary constant $\zeta^{lm}_1$ and $\zeta^{lm}_2$. Although our final results do not depend on these overall constants, we have obtained stable solutions for a range of values, $(\zeta^{lm}_1, \zeta^{lm}_2)\sim 10^{-3}-1$. Once the solution of $\chi^{lm}_{1(0)}$ and $\chi^{lm}_{2(0)}$ are obtained, we find out the solution of the remaining gauge invariant variables,  $\chi^{lm}_{3(0)}$ and $\chi^{lm}_{4(0)}$ using the following coupled equation,  
\be\label{eomchi34}
\bea
&f(r)\pr_r\chi^{lm}_{1(0)}+\chi^{lm}_{4(0)}=0,\\
&\pr_t \chi^{lm}_{1(0)}+f(r)\chi^{lm}_{3(0)}=0,
\eea
\ee
which are the governing equations (see Appendix \ref{schw.eom.tr} for the derivation) of the electromagnetic field for static Schwarzschild BH spacetime. Once all the zeroth-order solutions are obtained, we substitute these solutions in the source terms of the non-homogeneous eqs. \eqref{part.eq1} and \eqref{part.eq2} and again solve these differential equations numerically using ``StiffnessSwitching" method built in Mathematica.
\section{Setting up the boundary condition}\label{bc}
In the static Schwarzschild spacetime, asymptotic flatness allows us to normalize the scattering state solution of the matter field, so that the BH potential scatters an incoming wave with unit amplitude from infinity. Such normalization helps us to define \cite{Unruh:1976fm} the absorption cross section of a static BH. For our present case, however, the oscillatory nature of the ringing BH background does not result in a convenient asymptotic structure for the propagating field. To deal with such matters/fields with non-trivial asymptotic features, a hypothetical interaction/impact surface is sometimes utilized \cite{Hui:2019aqm} to study the transmission and reflection of waves in the BH spacetime. Including this surface helps one consistently define the asymptotic ingoing-outgoing modes, which otherwise would be very difficult to obtain. The physical arguments go as follows: by introducing the interaction surface, the spacetime is modified so that the asymptotic flatness can be exploited in the exterior region, and the solution in the interior region should be matched at the position of the interaction surface. One may immediately speculate as to whether the final results depend on the position of this interaction surface, as in the case in \cite{Hui:2019aqm}. We will address this point later. Nevertheless, to compute the absorption cross section for such a case, we proposed the methodology of introducing the hypothetical surface in our earlier paper \cite{Karmakar:2021ssg}, and for completeness, we outline the important steps here. 

\begin{figure}[t]
\centering
\includegraphics[scale=0.2]{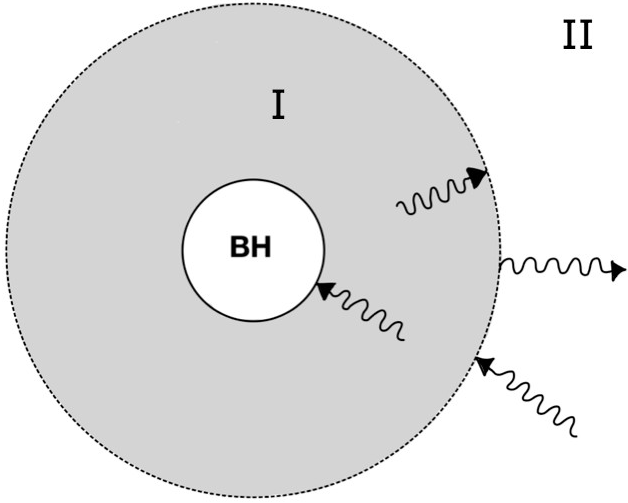}
\caption{We demonstrate the inclusion of the hypothetical interaction surface in the ringing black hole spacetime. Shaded region (${\bf I}$) is considered as the spacetime containing ringing fluctuations and outside region $({\bf II})$  is considered as the usual Schwarzschild spacetime.}\label{intsurface}
\end{figure}

We introduce the interaction surface, as shown in Fig.\ref{intsurface}, at some radial distance $r=r_{\rm int}$ from the BH event horizon. The wave coming towards the black hole is assumed to perceive the presence of a gravitational fluctuation once it hits the interaction surface.
 Outside the surface $r > r_{\rm int}$, the spacetime is assumed to be a static Schwarzschild BH, and this is the approximation which helps us to define the normalization of the incoming wave appropriately. However, the ringing fluctuation is notably present throughout the spacetime; therefore, the present approximation should be compensated by aggregating over the contribution of the individual position of the interaction surface and averaging with a large spatial domain. This methodology will be discussed in Sec.(\ref{meanACS}).
 Coming back, we label the region inside the interaction surface as region-I, and outside as region-II. With this setup, the individual mode (corresponding to $\km$) of the EM wave solution in region-I will assume, 
\be\label{tot_sol}
\bea
\chi_{j[\bf I]}^{\km lm}(t,r)&=e^{-\mi\km t}\chi^{lm}_{j(0)}(\km, r) + e^{-\mi {(\km+\omega)} t}\chi^{lm}_{j,{\rm p}}(\km , r)+e^{-\mi {(\km-\omega^*)} t}\bar\chi^{lm}_{j,{\rm p}}(\km, r)
\eea
\ee
where, $j\equiv (1,2)$, i.e. the expansion of both $\chi^{\km lm}_{1 [\rm I]}$ and $\chi^{\km lm}_{2[\rm I]}$ in the region-I. In region-II, it would be in the background of a static Schwarzschild BH, hence, the mode functions can be expressed as,
\be\label{region2}
\chi_{j[\bf II]}^{{\km lm}}(t,r)=\chi^{\km lm}_{j,0}(t,r) ~~\mbox{with} ~~{\cal L}_0(t,r)\chi^{\km lm}_{j,0}(t,r)= 0 .
\ee
\textit{We want to clarify that the notation used to denote the fields with subscript $\{j,0\}$ in the exterior of interaction surface in contrast to the same quantity with the subscription $\{j(0)\}/\{i(0)\}$ (see the discussion below \eqref{expand}) which represents the solution for the static Schwarzchild BH without the prior fluctuation}. The primary motivation for complicating the subscripts of the variable, $\chi$, rather than using different variables altogether, is to convey the fact that they represent the same physical quantity in various circumstances. 
Important to note that due to the interaction surface, the evolution of $\chi^{\km lm}_{j,0}$ in the exterior of the interaction surface will be different as compared to the situation of the usual static Schwarzschild background. Therefore, the interaction surface will naturally provide for the first two boundary conditions \eqref{bcs}. However, the above second order partial differential equation involving two variables $(t,r)$(the field solution at the interaction surface may not be separable in space and time) requires four boundary conditions. In region-II, the appropriate boundary condition would be as follows:  we provide two spatial initial conditions at the interaction surface $r_{int}$,
\be
\bea
\label{bcs}
&\chi^{\km{lm}}_{j[\bf II]}(t,r)|_{\forall{t},r\to{r_{\rm int}}}=\chi_{j[\bf I]}^{\km{lm}}(t,r)|_{\forall{t},r\to{r_{\rm int}}} ,\\
&\pr_{r}\chi_{j[\bf II]}^{\km{lm}}(t,r)|_{\forall{t},r\to{r_{\rm int}}}=\pr_r\chi_{j[\bf I]}^{\km{lm}}(t,r)|_{\forall{t},r\to{r_{\rm int}}} ,\\
\eea
\ee
and the remaining two conditions for the time we provide as boundary conditions at $t\to\infty$ as,
\be
\bea
&\chi_{j[\bf II]}^{\km{lm}}(t,r)|_{t\to{\infty},\forall{r}}=e^{-\mi\km t}{{\chi}}^{{\km}lm}_{j(0)}(r)|_{t\to{\infty},\forall{r}} ,\\
& \pr_t\chi_{j[\bf II]}^{\km{lm}}(t,r)|_{t\to{\infty},\forall{r}}=-\mi\km e^{-\mi\km t}{{\chi}}^{{\km}lm}_{j(0)}(r)|_{t\to{\infty},\forall{r}} .
\eea
\ee
Where the last two boundary conditions arise due to the following reason: at a fixed distance from the black hole $r$, ringing oscillation amplitude decays exponentially with time due to its quasinormal nature. Therefore, in $t\rightarrow \infty$ limit, the perturbative components of the gauge field, $(\chi^{\km lm}_{({\rm p})},\bar\chi^{\km lm}_{({\rm p})})$ which are the explicit function of $h_{\mu\nu}$ must also be vanishing at the interaction surface. 
Such a condition will naturally ensure the calculated absorption cross-section of ringing BH reducing to its static Schwarzschild value within the characteristic time scale of the oscillation $\tau \sim 2\pi/\omega$. 
By employing the boundary condition described in Eq.\ref{bcs}, we proceed to solve Eq.\ref{region2} in the domain, $([r_{\rm int}, \infty], [t_{\rm int}, \infty])$ lying inside light cone, $t\geq{r_*}$. 
For solving the partial differential equation, we have used the ``PDEDiscretization" method built in Mathematica. Whereas the convergence criteria for the solutions have been taken care of by setting the Accuracy goal ($\rm AG$) and Precision goal ($\rm PG$), which ensure that the solution, $\chi^{\km lm}_{j,0}(t,r)$, is obtained as far as the absolute error, ${\rm Er}<10^{-{\rm AG}}+|\chi^{\km lm}_{j,0}(t,r)|\times 10^{-\rm PG}$ \cite{Mathematica:url} is maintained. We have checked by varying ($\rm AG$,$\rm PG$) from $(8,8)$ to $(12,12)$ that the results remain stable with the specified error tolerance.

The asymptotic nature of the GWs is better expressed in the outgoing null coordinates. Therefore, after obtaining the solution, we perform a transformation into the outgoing null coordinate system. $(t,r)\to(u = t-r_*,r) $ using 
$\tilde{A}_{\mu}(x')=(\partial x^{\nu}/\partial x'^{\mu})A_{\nu}(x)$ \cite{Padmanabhan:2010zzb} and define the absorption cross-section accordingly as described in the following section.

\section{Defining Absorption Cross Section}\label{def.ACS}
In our framework, the procedure to determine the absorption cross section for an oscillating black hole follows directly from the method used in the case of static and stationary black holes. This is due to the fact that the exterior of the interaction surface, discussed above, is considered to be the static Schwarzschild black hole. Therefore, we first study the case of static and stationary black holes. To proceed, we write down the covariant conservation equation on a generic spacetime background with the associated Killing vector, $\xi^\mu$, as \cite{Padmanabhan:2010zzb}, 
\be
\nabla_\mu\left({\mathcal{T}^\mu}_\nu\xi^\nu\right)=0,
\ee
which can be explicitly checked using the covariant conservation of the symmetric stress energy tensor, $\nabla_\mu{\mathcal{T}^\mu}_\nu=0$, and the Killing equation $\nabla^\mu\xi^\nu+\nabla^\nu\xi^\mu=0$. Let us symbolize the quantity inside the parentheses of the above equation as the generalized four momentum, $P^{\mu}={\mathcal{T}^\mu}_\nu\xi^\nu$. To determine the conserved quantity, one needs to find out the allowed Killing vectors. In the case of the static spherically symmetric Schwarzschild black hole, the spacetime is described by the following line element, in $(u=t-r_*, r)$, 
\be\label{metric.ur}
ds^2=-f(r)du^2-2dudr+r^2(d\theta^2+\sin^2\theta d\phi^2). 
\ee
This metric allows the presence of time-like Killing vector, $\xi_{0} =\delta^\mu_0 \partial_{\mu}$. For this particular Killing vector, the associated conserved quantity turns out to be
\be
\mathcal{F}=\int{d^3x}\sqrt{-g_0(u,r)}P^0,
\ee
Where $g_0(u,r)$ is the determinant of the metric given in \eqref{metric.ur}. The above statement also implies $\pr_u \mathcal{F}=0$. Taking the time derivative on both sides of the above equation, we get
\be
\bea
\pr_u\mathcal{F}&=\int{d^3x}\pr_u\left(\sqrt{-g_0(u,r)}P^0\right)=-\int{d^3x}\pr_i\left(\sqrt{-g_0(u,r)}P^i\right).
\eea
\ee
By choosing a r-constant hypersurface and applying the divergence theorem in the above equation, one will arrive at
\be
\pr_u\mathcal{F}=-\int{r^2d\Omega}P^r\bigg|^{\infty}_{r_h}=-\int{r^2d\Omega}{\mathcal{T}^r}_u\xi^u\bigg|^{\infty}_{r_h}=-\left[\int{r^2d\Omega}{\mathcal{T}^r}_u\bigg|_{r\to\infty}-\int{r^2d\Omega}{\mathcal{T}^r}_u\bigg|_{r_h}\right].\\
\ee
Where, $d\Omega\equiv \sin\theta d\theta d\phi$. We can see that two terms in the last equality of the above equation are equal by the fact that $\pr_u\mathcal{F}=0$, as mentioned previously. Therefore, one may infer that the energy being absorbed by the black hole horizon per unit time could be equated with $\pr_u\mathcal{F}\big|_{r\to\infty}$. Now, the definition of the absorption cross section \cite{Unruh:1976fm} is the amount of energy being absorbed by the black hole horizon divided by the incident energy density. We define the incident energy density as, $\pr_u \mathcal{G}={\mathcal{T}^z}_u$, which sometimes called as energy density current \cite{Cardoso:2019dte}. Utilizing this methodology we obtain the absorption cross section, for individual modes, in terms of the stress-energy tensor of the fields as \cite{Cardoso:2019dte},
\be\label{def1.abs}
\sigma^{\km l}_{\rm ring}(u,r_{\rm int})\equiv\frac{\pr_u\mathcal{F}^{\km{l}}}{\pr_u\mathcal{G}^{\km}}=\frac{\int{r^2d\Omega}{\mathcal{T}^r}_u}{{\mathcal{T}^z}_u}\bigg|_{r\to\infty}.
\ee
Note that the index, $m$, associated with azimuthal mode does not appear here for reasons that will be clarified in the following discussion.
Nevertheless, the above expression, in principle, should be evaluated at spatial infinity (one may also look at eq.(23) of \cite{Crispino:2008zz} for a similar definition using current density). For practical purposes, we have presented our results considering $r=75r_h$, however, we have checked that all the results are stable with $r=50 r_h-100 r_h$. Important to note that, asymptotically, both the numerator and denominator become independent of the spatial coordinate and only depend on the time coordinate, u. However, the dependence on the $r_{\rm int}$ will naturally enter through the EM fields with the initial condition set at the interaction surface. We will now explicitly express the numerator and denominator of \eqref{def1.abs} in terms of EM field \eqref{modedecom}, using the gauge-invariant variables from \eqref{invvar}. Of course, one has to take care of the transformation from ($t,r$), to ($u,r$) (see the discussion in Appendix.\ref{eomucoord}), which will attribute a ``tilde" sign over the field- variables. Following these steps the numerator of \eqref{def1.abs} can be expressed in the following manner after integrating over the spherical surface at $r\to\infty$ (see Appendix.\ref{derv.numer.abs} for detailed derivation),
\be\label{numerator.def.abs}
\bea
\pr_u\mathcal{F}^{\km l}&=l(l+1)\Big[\frac{1}{2}\Big\{\tilde{\chi}^{\km lm}_{3,0}(u,r)\tilde{\chi}^{\km lm^*}_{4,0}(u,r)+\tilde{\chi}^{\km lm}_{4,0}(u,r)\tilde{\chi}^{\km lm^*}_{3,0}(u,r)\\
&~~~~~~~~~~~~~+\pr_r \tilde{\chi}^{\km lm}_{2,0}(u,r)\pr_u \tilde{\chi}^{\km lm^*}_{2,0}(u,r)+\pr_u \tilde{\chi}^{\km lm}_{2,0}(u,r)\pr_r \tilde{\chi}^{\km lm^*}_{2,0}(u,r)\Big\}\\
&~~~~~~~~~~~~-\Big\{\tilde{\chi}^{\km lm}_{4,0}(u,r)\tilde{\chi}^{\km lm^*}_{4,0}(u,r)+\pr_u \tilde{\chi}^{\km lm}_{2,0}(u,r)\pr_u \tilde{\chi}^{\km lm^*}_{2,0}(u,r)\Big\}\Big].\\
\eea
\ee
Coming to the evaluation of the above expression, we find out the solution of  $\tilde{\chi}^{\km lm}_{1,0}(u,r)$ and $\tilde{\chi}^{\km lm}_{2,0}(u,r)$ by transforming the solution obtained in $(t,r)$ \eqref{region2}, whereas for the solution of the remaining gauge invariant variables, we use the following coupled equation,  
\be
\bea
&\tilde{\chi}^{lm}_{3,0}(u,r)=-\pr_r\tilde{\chi}^{lm}_{1,0}(u,r)\\
&\tilde{\chi}^{lm}_{4,0}(u,r)=\pr_u\tilde{\chi}^{lm}_{1,0}(u,r)+\tilde{\chi}^{lm}_{3,0}(u,r),
\eea
\ee
which are the governing equations (see appendix.\ref{eomucoord} for the derivation) of the electromagnetic field for the static Schwarzschild BH spacetime in (u,r) coordinates. The procedure outlined above supports our earlier assertion that there's no need to directly solve all the gauge invariant variables. 

Now, the incident energy density per unit time for a plane EM wave propagating in the z-direction is constructed as
\be
\pr_u\mathcal{G}^{\km}={\mathcal{T}^z}_u .
\ee
Having evaluated the solution of the EM wave in spherical coordinates with a damped oscillating time-dependent feature we find it convenient to recast the expression for incident energy density in the spherical coordinates as (one may look at \eqref{fluxzu.in} and \eqref{fluxsu.in}), 
\be\label{mainfluxsu.in}
\bea
\pr_u\mathcal{G}^{\km}&=\Big\{\frac{1}{2}g^{ss}_0(u,r)(\pr_u A_s-\pr_s A_u)(\pr_r A^*_s-\pr_s A^*_r)+c.c.\Big\}\\
&~~~~~~~~~~~~~~~~~-g^{ss}_0(u,r)(\pr_u A_s-\pr_s A_u)(\pr_u A^*_s-\pr_s A^*_u) .
\eea
\ee
Once again $``s"$ represents coordinates, ($\theta,\phi$), as before. Also, notice that all the raising and lowering of indices have been done with respect to $g^{\mu\nu}_0(u,r)$, keeping in mind that the incident energy density should be evaluated at a considerable distance from the interaction surface, i.e. in the exterior region where spacetime is characterized as Schwarzschild.
With the help of \eqref{modedecom} and \eqref{invvar} one can express (of course, one has to take care of the coordinate transformation from $(t,r)$ to $(u,r)$) the field combinations of \eqref{mainfluxsu.in} in terms of the invariant variables as,
\be\label{asymcomb1}
\bea
&\pr_u A_s-\pr_s A_u=\sum_{lm}\left[\left(\pr_r\tilde{\chi}^{\km lm}_{1,0}-\pr_u\tilde{\chi}^{\km lm}_{1,0}\right)\Psi^{lm}_s(\Omega)+\pr_u\tilde{\chi}^{\km lm}_{2,0}\Phi^{lm}_s(\Omega)\right],\\
&\pr_r A_s-\pr_s A_r=\sum_{lm}\left[\pr_r\tilde{\chi}^{\km lm}_{1,0}\Psi_s+\pr_r\tilde{\chi}^{\km lm}_{2,0}\Phi_s\right].\\
\eea
\ee
The remaining task is to figure out the incident part of the EM wave, $\tilde{\chi}^{\km lm}_{1,0}$ and $\tilde{\chi}^{\km lm}_{2,0}$ that can be obtained by assuming an approximate asymptotic form,
\be\label{asymp.soln}
\bea
& \tilde{\chi}^{\km lm}_{1,0}(u,r)=\mathcal{N}^{\km lm}_1[\mathcal{I}_1(u)e^{-\mi\km(u+2r_*)}+\mathcal{R}_1(u)e^{-\mi\km u}], \\ 
&\tilde{\chi}^{\km lm}_{2,0}(u,r)=\mathcal{N}^{\km lm}_2[\mathcal{I}_2(u)e^{-\mi\km(u+2r_*)}+\mathcal{R}_2(u)e^{-\mi\km u}] ,\\
\eea
\ee
where the coefficients, $(\mathcal{I}_1(u),\mathcal{R}_1(u))$ and $(\mathcal{I}_2(u),\mathcal{R}_2(u))$ are related to ingoing (incident) and outgoing (reflected) wave amplitude of $\tilde{\chi}^{lm}_{1,0}(u,r)$ and $\tilde{\chi}^{lm}_{2,0}(u,r)$ respectively. We evaluate these amplitudes numerically from the solutions of $\tilde{\chi}^{lm}_{1,0}(u,r)$ and $\tilde{\chi}^{lm}_{2,0}(u,r)$. Now comparing the above two expressions with the same quantities constructed (see Appendix.\ref{normfac}) for circularly polarized incident plane EM wave we derive the normalization factors as, 
\be
\bea
&\mathcal{N}^{\km lm}_1=-\mi(-1)^{l+1}\delta_{m1}\sqrt{\frac{4\pi(2l+1)}{l(l+1)}}\frac{1}{2\km \mathcal{I}_1(u\to\infty)} ,\\
&\mathcal{N}^{\km lm}_2=(-1)^l\delta_{m1}\sqrt{\frac{4\pi(2l+1)}{l(l+1)}}\frac{1}{2\km\mathcal{I}_2(u\to \infty)} .
\eea
\ee
Where $\delta_{m1}$ appears due to the fact that the EM field potentials, $A_\mu(x)$, transforms as a vector, and that shows up in the spherical wave expansion of the incident circularly plane polarized EM wave (see details in the appendix \ref{normfac}) coming from z-direction, with azimuthal index $m=1$ \cite{Crispino:2008zz} similar to that of scalar field analysis where $\delta_{m0}$ appears due its spin zero property\cite{Cardoso:2019dte, Karmakar:2021ssg}. We also normalize the combinations of \eqref{asymcomb1} by dividing the right hand side with $\sum_{lm}(-1)^{l+1}\delta_{m1}\sqrt{\frac{4\pi(2l+1)}{l(l+1)}}[\Phi^{lm}_s(\Omega)+\mi\Psi^{lm}_s(\Omega)]$, consequently the gauge invariant combinations finally become,  
\be\label{asymcomb2}
\bea
&\pr_u A_s-\pr_s A_u=\xi_{us}(\km,u)(-\mi\km A'_s(u,\textbf{r})) ,\\
&\pr_r A_s-\pr_s A_r=\xi_{rs}(\km,u)(-2\mi\km A'_s(u,\textbf{r})),
\eea
\ee
with the nontrivial time-dependent coefficients, $\xi_{us}(\km,u), \xi_{rs}(\km,u)$ with the property that they become unity as the static limit, $u\to \infty$, of the ringing BH is approached (for details one may look at Appendix.\ref{normfac}). Importantly, the sum in \eqref{asymcomb1}with normalization factors exhibits a convergent nature for increasing $l$ having alternative $+/-$ sign in the summation, therefore, we have taken up to $l=6$ for our numerical analysis. Whereas, $A'_s(u,\textbf{r})$, denotes the spherical polar expansion \cite{Crispino:2008zz} of the incident circularly polarized EM wave, which can be suitably factored out from the expression of the incident energy density. 
\begin{figure*}[!htb]
\includegraphics[width=\linewidth]{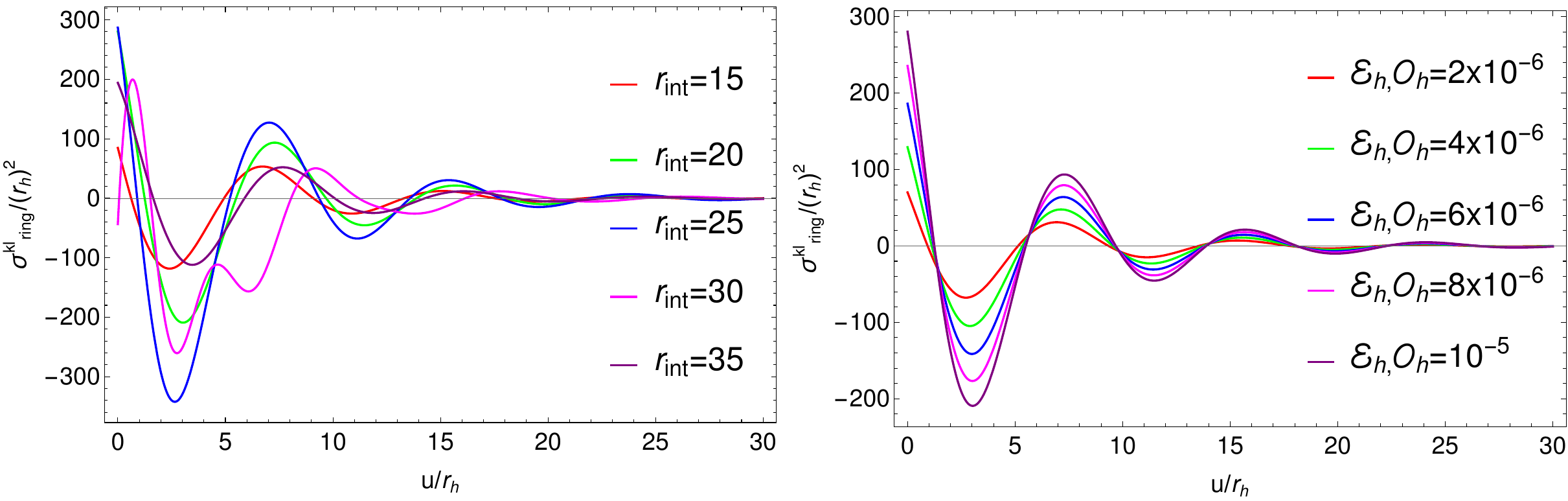}
\caption{On the left panel absorption cross section of the ringing BH for the EM field has been plotted with time, $u$, for various interaction surfaces, $r_{int}$, considering frequency, $\km=0.1 r^{-1}_h$ and $l=1$. On the right panel, we have plotted the same by varying the background amplitude, $\mathcal{E}_h, \mathcal{O}_h$, considering frequency, $\km=0.1 r^{-1}_h$ and $l=1$ at a particular interaction surface $r_{int}=20 r_h$. All the parameters written inside the plots are in units of $r_h$.}\label{diffrintdiffamp}
\end{figure*}
\section{Numerical Results}\label{num.results}
As previously mentioned, the ringing BH background is constructed taking only the quadrupole oscillation \cite{Rezzolla:2003ua} or the lowest $l(=2)$ mode with the corresponding quasinormal frequency $\omega = (0.74734-\mi ~0.17792)(r_h)^{-1} $ \cite{Chandrasekhar:1975zzb, Molina:2010fb}. $r_h=2M$ is the horizon radius for the  Schwarzschild BH. The chosen frequency is known to be the long-lived one among all the other modes. Further, the frequency for both even and odd parity perturbation was found to be the same as per the existing literature \cite{Chandrasekhar:1975zza, Chandrasekhar:1975zzb, Molina:2010fb}. 
All physical quantities and associated parameters have been made dimensionless utilizing the characteristic scale, $r_h$.
We consider the solution of the background ringing field for which the perturbative scheme is valid ensuring $\delta g/g_0 \propto h^\mu_\mu \ll 1 $ for a diverse range of initial parameters. Once the solution for the ringing black hole background is obtained, we solve for the EM field Eq.\ref{region2}. Unless stated throughout our presentation we have shown the results for a fixed background amplitude, $\mathcal{E}_h, \mathcal{O}_h\sim 10^{-5}$ within the perturbative limit. As mentioned earlier, as a consistency check we also reproduce the well studied static limit of the absorption cross-section \cite{Crispino:2007qw} of Schwarzschild BH for EM field in the limit, $\lim_{u\rightarrow \infty}\sigma^{\km l}_{\rm ring}(u,r_{\rm int}) = \sigma^0(\km,l)$. 

\begin{figure*}[!htb]
\includegraphics[width=\linewidth]{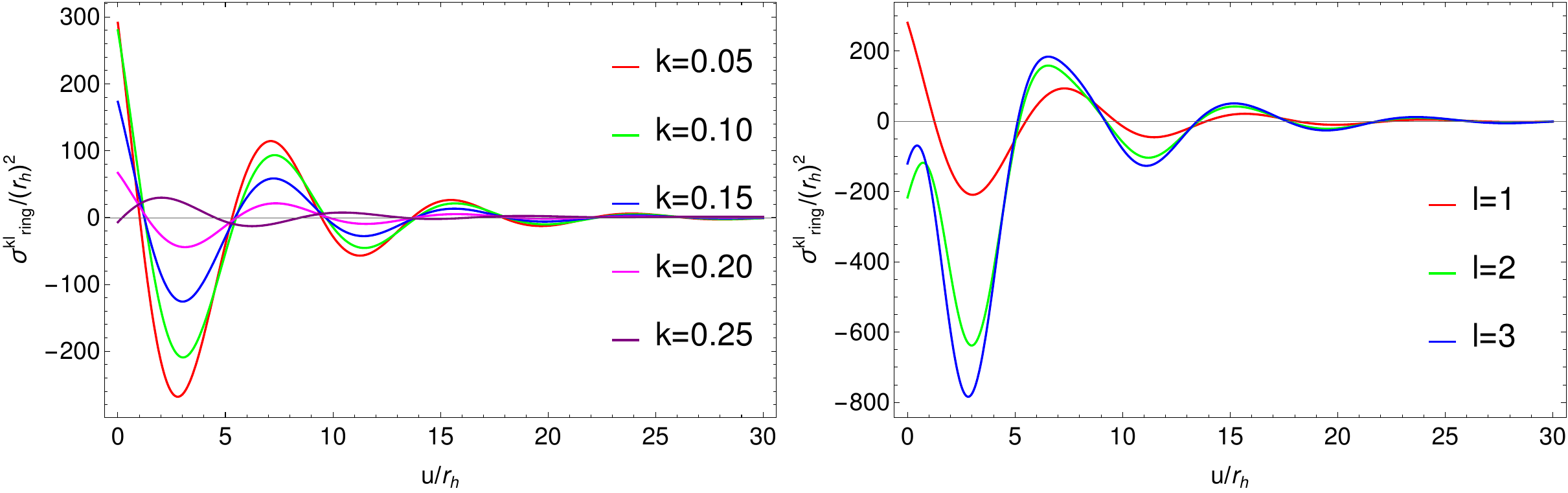}
\caption{On the left panel absorption cross section of the ringing black hole for the EM field has been plotted with time, $u$, by varying the frequency, $\km$, of the EM field considering $l=1$. On the right panel, we have plotted the same various angular modes, $l$, of the EM field considering $\km=0.1 r^{-1}_h$. All the parameters written inside the plots are in units of $r_h$.}\label{diffkdiffl}
\end{figure*}

We evaluate the absorption cross section at $r\sim 75 r_h$ and further numerically ensure that all the necessary results remain intact even afterwards, apart from very small fluctuations induced by numerical precisions due to the extension of the domain of the solutions. Our final results are summarized in Figs.\ref{diffrintdiffamp} and \ref{diffkdiffl}. 

According to our construction, time-dependent boundary conditions at the interaction surface would introduce time-varying features in the absorption cross section. Such a transient nature of the GWs is clearly seen to be imprinted in the behaviour of EM absorption cross section and exhibits the quasi-periodic oscillation along with its characteristic time scale. One can indeed observe that within the GW time scale $\tau_{\rm GW} \sim 35 r_h$ associated with the quasinormal frequency $\omega = (0.74734-\mi ~0.17792)(r_h)^{-1} $, the EM absorption cross section $\sigma^{\km l}(u,r_{\rm int})$ undergoes a time-dependent oscillation, which can assume large negative values, and this negative amplitude of the absorption cross section precisely signifies the phenomena of superradiant scattering.

We identify five main theory parameters of our interest $(\km, l, r_{\rm int})$ and the GW amplitudes $({\cal E}_h, {\cal O}_h)$. However, for all our practical purposes, we choose both the GW amplitudes to be the same. With decreasing GW amplitude the EM superradiant amplitude is expected to decrease which we have consistently observed in our numerical results, and indeed be seen in the second panel of the Fig. \ref{diffrintdiffamp}. This also provides us confidence as a consistency check of our numerical methodology. 

\textit{Interaction surface:} We provide the plots choosing the interaction surface located at $r_{\rm int}\sim 20 r_h$. Interestingly, the superradiant amplitude turned out to be maximum at $r_{\rm int}\sim 20-25 r_h$ for all angular modes, $l$ and frequency, $\km$ within our consideration. For example $\sigma\sim -354 r^2_h$ (fig.\ref{diffrintdiffamp}) is the maximum value for $l=1$ at frequency $\km \sim 0.1 r^{-1}_h$ and $r_{\rm int}\sim 25 r_h$. The reason behind this effect has been argued in the following discussion. As it turns out the final results are significantly dependent on $r_{\rm int}$. However, this result does not represent the outcome as identified by an asymptotic observer. The matter field, being coupled with the ringing black hole background, interaction with spacetime should be taken throughout the region from the horizon to the spatial infinity. 
Hence, to obtain the physically sensible result we have to integrate out the $r_{\rm int}$ that will be discussed in the next section. Nevertheless, the characteristics exhibited by the absorption cross section for the individual positions, such as, for $r_{\rm int}$ corresponding to the maximum enhancement, bear the features of the aggregated result. Importantly, the time profile of the absorption cross section for different $r_{\rm int}$ suggests that the aggregated effect may not vanish.

In the perturbative framework, the scatter solution of a particular order depends on the GW background and lower-order EM solution as the source. Particularly when the maximum amplitude of the lower order EM solution appears away from the BH, GW amplitude is always maximum near the horizon.  
The resultant of those two contributing factors effectively decides the location of the interaction surface for which superradiant amplitude is maximized. This is reminiscent of much studied \cite{Brito:2015oca, Dai:2023ewf} superradiance from rotating BHs whose effective potential maximizes itself a little away from the event horizon.

In fig.\ref{diffkdiffl} we have shown the behaviour of the superradiant absorption cross section with different frequencies. Like the usual case of static charged or rotating BHs \cite{Leite:2018mon}, a ringing BH also exhibits superradiance at low frequency with distinct time-varying features induced by the GWs. We have obtained the expected cut-off frequency beyond which the superradiance ceases to exist for different angular momentum $(l)$. For example, we obtain $\km_{\rm max}=0.28 r^{-1}_h$ for $l=1$ above which superradiance vanishes, and similarly for $l=2, \km_{\rm max}\sim 0.55 r^{-1}_h$, and $l=3, \km_{\rm max}\sim 0.8r^{-1}_h$. A non-trivial point to note is that for the higher angular momentum mode the cut-off frequency $\km_{\rm max}$ increases contrary to the usual expectation as exciting higher frequency mode would be difficult.  
For example, only long wavelength modes are amplified in the usual superradiance phenomena from black hole \cite{Starobinskil:1974nkd,Crispino:2007qw, Leite:2017zyb, Crispino:2008zz,Leite:2018mon}. For the ringing case, however, the reason could be complicated mode coupling and their competition in the source terms (see details in the  Appendix.\ref{source}). Our result suggests that the perturbative scheme may not be valid above certain angular modes which could be further inferred from the increasing superradiant amplitude for higher angular mode $l$ shown in Fig.\ref{diffkdiffl}. However, this is not a unique feature to the ringing case, for the moving BH system, similar behaviour in the absorption cross-section has been observed as discussed in \cite{Cardoso:2019dte}. Nonetheless, 
further details need to be investigated to decode such behaviour, and that is beyond the scope of our present study.  

\section{Mean absorption cross section and Energy extraction at spatial infinity}\label{meanACS}
\begin{figure}[t]
\centering
\includegraphics[scale=0.43]{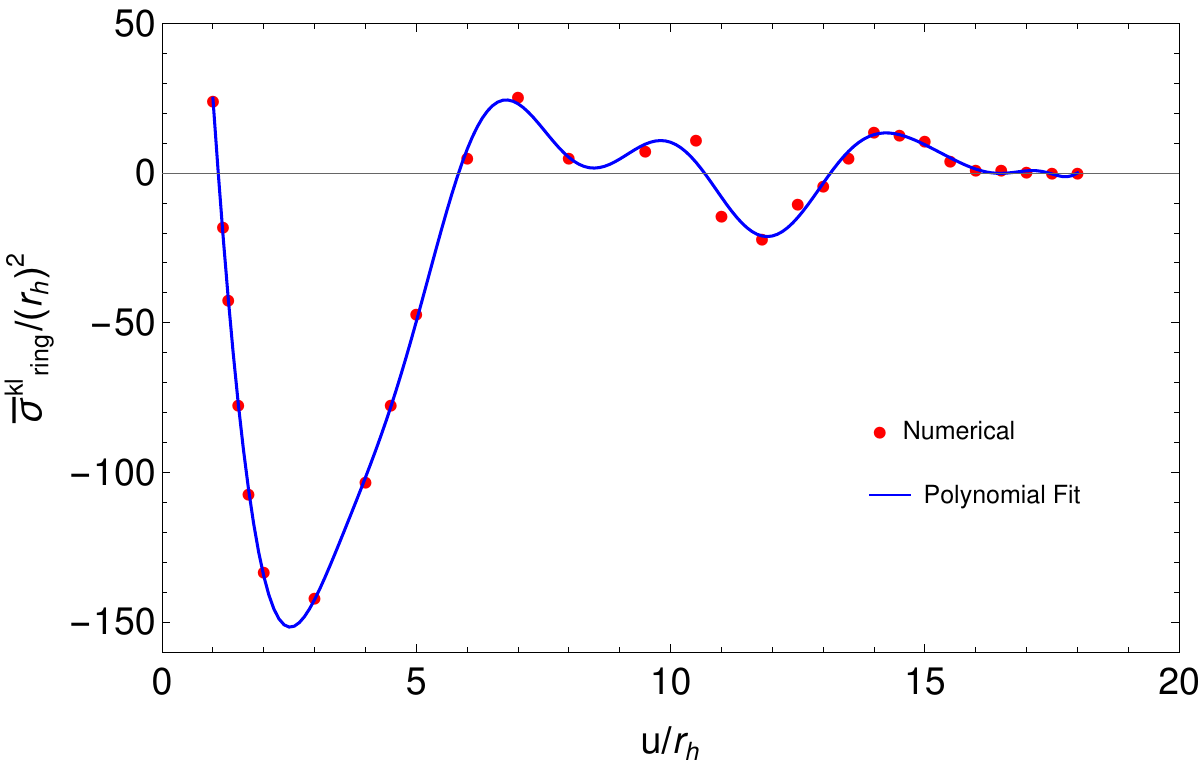}
\caption{Mean absorption cross section has been plotted with time by averaging over the position of the interaction surface, $r_{\rm int}$ for a fixed frequency, $\km=0.1 r^{-1}_h$ and $l=1$.}\label{Integrated}
\end{figure}

To find out the absorption cross section in a time-dependent background, such as that of a ringing black hole, we introduced the concept of the interaction surface. As a computational tool, the position of the interaction surface, $r_{\rm int}$ should not turn up in the final result. Although the reader has already noticed that the absorption cross section significantly depends on $r_{\rm int}$, this does not provide the complete picture. This is because the incoming wave, in principle, interacts with the gravitational background at every point. Therefore, we further compute the mean value of the absorption cross section, which is independent of the position of the interaction surface, $r_{\rm int}$, by integrating over $r_{\rm int}$, and dividing by a large spatial domain in the following manner,
\be
\bar{\sigma}^{\km l}_{\rm ring}(u)=\frac{\int\sigma^{\km l}_{\rm ring}(u,r_{\rm int})dr_{\rm int}}{\int dr_{\rm int}}.
\ee
Although this definition, a priori, may seem arbitrary, it correctly reproduces the standard result for the absorption cross section of a static Schwarzschild black hole in the expected limit as $u\to \infty$. To numerically evaluate the above quantity, first, we have obtained the absorption cross section at a particular instant of time for $r_{\rm int}\in (5r_h,50r_h)$. In this interval, we have fitted the points using a Polynomial fit, which will provide a functional form of the absorption cross section in terms of the $r_{\rm int}$ and integrated it in the same interval. Repeating this procedure for various instances of time, we have generated the plot of Fig.\ref{Integrated}. One may extend the domain of $r_{\rm int}$, but that will negligibly affect the result, as we have already seen in Fig.\ref{diffrintdiffamp} that the maximum amplitude occurs at $r_{\rm int}=25r_{h}$ and decreases on both sides. The integration interval for $r_{\rm int}$ is selected such that extending it further (outside the black hole) does not alter the magnitude or time profile, while also reducing computation time. The mean absorption cross section presented in Fig.\ref{Integrated}, thus independent of $r_{\rm int}$, however, shares very much the same feature as given in Fig.\ref{diffrintdiffamp} and Fig.\ref{diffkdiffl}.

{\textbf{Energy Extraction:}} By definition, the absorption cross section entails the fractional energy gain (see \cite{Teukolsky:1974yv}), if it happens so, as can be realized by looking at the numerator in the definition \eqref{def1.abs}, expanding which in terms of the fields, we obtain,
\be\label{energy.extract}
\bea
\int{r^2d\Omega}{\mathcal{T}^r}_u&\sim\Big[\left|\mathcal{N}^{\km lm}_1\right|^2\left(|\mathcal{I}_1(u)|^2-|\mathcal{R}_1(u)|^2\right)+\left|\mathcal{N}^{\km lm}_2\right|^2\left(|\mathcal{I}_2(u)|^2-|\mathcal{R}_2(u)|^2\right)\Big],\\ 
&\sim(\pr_u E^{tot}_{\rm in}-\pr_u E^{tot}_{\rm out}),
\eea
\ee
where $E_{\rm in}$ and $E_{\rm out}$ represent respectively the total ingoing and outgoing energy flux, which can be obtained by substituting \eqref{asymp.soln} in the numerator of \eqref{def1.abs}. Note that there will be additional terms in the first line of the above equation; however, the contribution of them turns out to be negligibly small. Nevertheless, the negative values of the absorption cross section thus imply, $\pr_u E^{tot}_{\rm in}<\pr_u E^{tot}_{\rm out}$, which happens in certain time intervals as shown in Fig.\ref{diffrintdiffamp},\ref{diffkdiffl} and Fig.\ref{Integrated}. In these durations, the electromagnetic field will extract energy from the ringing BH. Of course, the positive part of the absorption cross section, presented in Fig.\ref{diffrintdiffamp},\ref{diffkdiffl} and Fig.\ref{Integrated} suggest that there will be the absorption of radiation in some intervals. However, the superradiance will happen in the rest of the intervals within the decay time scale of the ringdown phase. An observer sitting at spatial infinity will receive this amplified radiation as a repeated flash, which eventually decays, exhibiting the transient nature of the ringing phase of the BH.
\section{Observability of the superradiant signal and primordial black hole merging}
\begin{figure}[t]
\centering
\includegraphics[scale=0.7]{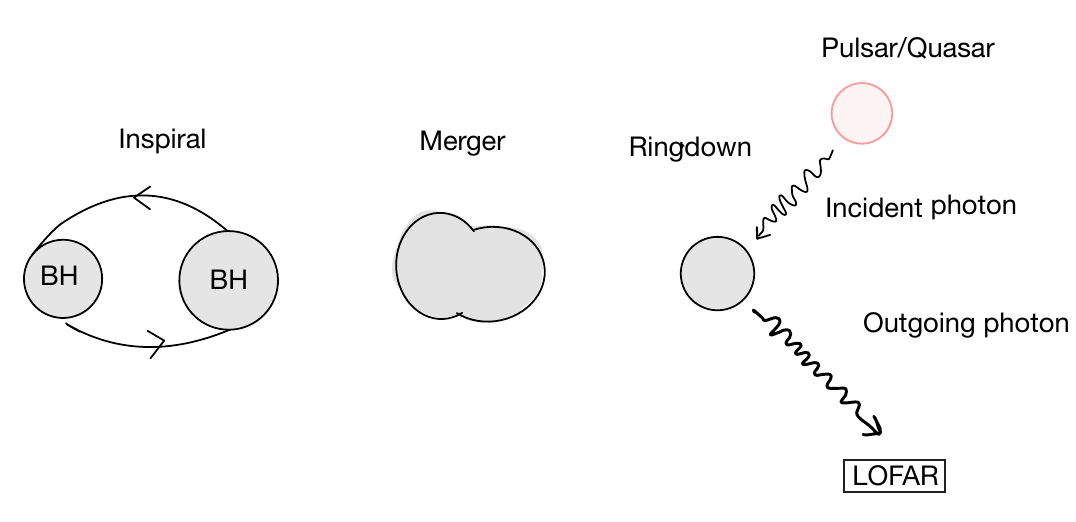}
\caption{Schematic diagram of the BH merging phase, illustrating a scenario where the BH masses and initial configuration are chosen such that the quiescence time occurs in the current era of the universe. An incoming photon from a Pulsar or Quasar, intended as a target source for LOFAR, may scatter off the ringing black hole and undergo amplification.}\label{Pulsar_photon}
\end{figure}
In the last decade, the GW observations by LVK \cite{LIGOScientific:2016aoc, LIGOScientific:2016emj, LIGOScientific:2016vbw, LIGOScientific:2016vlm, LIGOScientific:2016sjg, LIGOScientific:2017ycc, LIGOScientific:2017vwq, LIGOScientific:2020stg, LIGOScientific:2016jlg, LIGOScientific:2020aai, LIGOScientific:2020zkf, LIGOScientific:2017vox, LIGOScientific:2023fpk, KAGRA:2023pio, KAGRA:2022twx} have demonstrated that the BHs have extended mass function. On the other hand, BHs, which are supposed to have been formed due to large density fluctuation in the early universe, span a wide range of mass. Therefore, these BHs, classified as primordial black holes (PBHs), could be viable candidates for the source of GWs as detected by LVK, given that the PBHs undergo the inspiral-merger-ringdown process.   

PBHs are assumed to form due to gravitational collapse during the usual cosmological evolution if the density perturbation is very large in some spatial region (details of the formation mechanism can be found in \cite{Carr:2018nkm, Carr:2020gox, Escriva:2022duf, Sasaki:2018dmp} and for an observational perspective \cite{Carr:2023tpt, Kashlinsky:2019kac}). If the universe is radiation-dominated, and the density perturbation satisfies the appropriate condition, the PBH formation mass can be estimated from \cite{Sasaki:2018dmp} 
\begin{equation}
M\sim M^2_{\rm pl} H^{-1}
\end{equation}
where the Hubble radius $H=\sqrt{\rho_r/3M^2_{\rm pl}}$, $M_{\rm pl}$ is the reduced Planck mass. Whereas, radiation energy density is given by $\rho_r=(\pi^2g_*/30) T^4$, with $T$ being the radiation temperature and $g_*$ signifying the effective number of relativistic degrees of freedom. Utilizing these expressions we can estimate the early universe temperature when a BH of particular mass is formed, $T^2\sim M^3_{\rm pl}\sqrt{90/(\pi^2g_*)}/M$. For example the BH of mass $M\sim 10^{-2}M_{\odot}$ at present is formed when the temperature of the universe was $T\sim 128$ GeV. If a substantial number of such primordial black holes (PBHs) formed during the early universe, their mergers could enter the ringing phase, resulting in superradiant amplification at frequencies within the detectable range, as discussed below. Moreover, it can be shown that if the PBHs mass $M \lesssim 10^{15}$ gram, those will evaporate by now given the age of the universe $\sim 10^{10} yr$ \cite{Page:1976df, Hawking:1975vcx, Baldes:2020nuv}. Therefore, in principle, one can consider the mass range of the PBH within $M>10^{-18} M_{\odot}$, that can survive till now and merge. In the following discussion, we will explore this mass range to put forward the probable scenario of detecting superradiantly enhanced photon flux, due to the ringing PBHs, in the ground-based telescopes.

The possibility of detecting the flux of EM waves with enhanced amplitude in ground-based telescopes depends on the operational frequency ranges of these observatories. In terms of the dimensionless cutoff frequency, $\km_{\rm max}$, introduced before, the cutoff frequency for the occurrence of superradiance can be expressed in the real unit as, $\omega\sim 10^5 \km_{\rm max}\left({M_{\odot}}/{M}\right)\SI{}{Hz}$, where, $M_{\odot}$ is the solar mass and $M$ is the BH mass under consideration. As an example for a solar mass BH, with EM field parameters $l=1$, $\km_{\rm max}=0.28$, we obtain $\omega\sim 10^4\SI{}{Hz}$, which is in the very low, radio frequency range. Below this frequency, we would expect to observe the superradiant scattering. 
However, the existing ground-based observatories, such as LOFAR (Low-Frequency Array), are sensitive to the EM wave within the frequency band, $10-240 \SI{}{\mega\hertz}$, or wavelength $\sim 30-1.2 m$ \cite{vanHaarlem:2013dsa, Corstanje:2016mia}. Converting this operational range of frequency in our present context, we obtain the required BH mass as $M< 10^{-2}\sim 10^{-1}M_{\odot}$, from which superradiant scattering of EM field could be relevant for detection. However, the smallest compact object in the binary coalescence found by LIGO, VIRGO \cite{LIGOScientific:2020zkf} is $M\sim 2.6 M_{\odot}$, which falls in the mass gap region of a neutron star and very light black hole. Further, for the astrophysical BHs, the lower mass limit is set by Chandrasekhar limit $M \sim1.4 M_{\odot}$ \cite{Chandrasekhar, Hawking:1987en}. Therefore, to be able to observe any superradiant scattering of the EM field, the required mass of the BHs should not be of astrophysical origin. The interesting way out appears to be primordial BHs (PBHs) which have garnered a lot of interest in the recent past, particularly in cosmology \cite{RiajulHaque:2023cqe, Musco:2018rwt}. At this point, we should mention that there are some interesting studies already \cite{Calza:2023rjt, Branco:2023frw, Rosa:2017ury} on superradiance from PBHs. 

Our analysis suggests that if PBHs undergo the inspiral-merger-ringdown process, they should exhibit superradiance phenomena during the ringdown phase, as indicated by our analysis. Moreover, depending upon their merger rate and distribution, they can lead to superradiant scattering within a wide observable frequency range $10^6~\SI{}{\hertz}<\omega<10^{22}~\SI{}{\hertz}$. However, important to note that the cross-section being $\sigma\propto r^2_h\propto M^2$, lowering the mass naturally reduces the superradiant signal strength. Therefore, PBH mass within the range $M\sim 10^{-1}M_{\odot}-10^{-2}M_{\odot}$ could be of importance from the observational point of view. Let us now discuss the coalescence time for PBHs, whose remnants fall within this mass range. For a simplified estimation, we assume that the merger is driven solely by gravitational wave emission. In the non-relativistic regime, Peters' formula \cite{Peters:1964zz} gives the merger time as,
\be
t_{\rm merger}=\frac{3}{85}\frac{c^5}{G^3}\frac{a_0^4}{M_1M_2(M_1+M_2)}(1-e_0^2)^{\frac{7}{2}},
\ee
where $M_1$ and $M_2$ represent the masses of the PBH binary companions. Whereas $a_0$ and $e_0$ are the initial separation and eccentricity of the PBH binary, respectively. For a PBH binary with $M_1\sim 0.003 M_{\odot}$ and $M_2\sim 0.007 M_{\odot}$ , considering initial separation, $a_0\sim 0.01 {\rm Au}$ and eccentricity $e_0\sim 0.99$, the above expression yields a merger time of $t_{\rm merger}\sim 10^{10}~yr$. As per this estimate, PBHs formed during the radiation-dominated era of the early universe with such a configuration would be merging in the present epoch. Therefore, photons emitted by various astrophysical events in the present time, such as pulsars or quasars, may scatter off the ringing black hole and undergo amplification. In Fig.\ref{Pulsar_photon}, we have provided a schematic diagram of the interaction between the photon and the ringing BH.

Several ground-based radio observatories are fully operational to explore EM signals, such as the Giant Metrewave Radio Telescope (GMRT) with frequency range, $\omega\gtrsim 50 {\rm M}\SI{}{\hertz}$ \cite{gmrttifr, Gupta2017, Intema:2016jhx}, Square Kilometer Array (SKA)\cite{Weltman:2018zrl}, and Very Large Array (VLA) \cite{VLA} other than LOFAR (already mentioned). Moreover, the latest addition, LOTAAS (LOFAR Tied-Array All-Sky Survey)\cite{Sanidas:2019stw, Michilli:2018sqb, Michilli:2019uzx, Tan:2020alu} is capable of receiving signals throughout the northern hemisphere.
\begin{table}[t]
 \centering
\begin{tabular}{|c|c|c|} 
 \hline
 Observatory  & Frequency range & Precision time \\
 \hline\hline
 LOFAR & $10-240\SI{}{\mega\hertz}$ & $\sim 0.1 ns$  \\
\hline
 GMRT & $ 50 \SI{}{\mega\hertz}-1.5 \SI{}{\giga\hertz}$ &$\sim 81\mu s$ \\ 
 \hline
 VLA & $74\SI{}{\mega\hertz}$ - $50\SI{}{\giga\hertz}$ & $\sim 5 ms$ \\
 \hline\hline
BH mass range & Superradiant & $\sigma^{kl}_{\rm ring}$\\
 & frequency & time scale $(\tau)$\\
 \hline
$ 0.1 M_{\odot}-0.01 M_{\odot}$ &  $ 1 \SI{}{\mega\hertz}-10 \SI{}{\mega\hertz}$ & $ 20\mu s-2 \mu s$\\
\hline
\end{tabular}
\caption{The frequency and precision-time-scale of existing observatories, along with the same quantities corresponding to EM absorption cross-section for a particular range of detectable BH mass range}
    \label{tabobs}
\end{table}
To this end, we must point out the time scale of the evolving superradiant amplitude  $\tau \sim 200 (M/M_{\odot}) \mu s$, which is of the order of the GW oscillation time scale. Therefore, to measure such a signal, the sensitivity in the time measurement is extremely important, and in this range with $M\sim 10^{-2} M_\odot$, implying $\tau \sim 2\mu s$, only LOFAR with precision $\sim 0.1$ ns \cite{Corstanje:2016mia}, could detect those. To put the above estimates in perspective, in the table.\ref{tabobs}, we provide the frequency range and precision time scale of the existing observatories along with the frequency ranges corresponding to the superradiant absorption cross-section and its oscillation time scale, $(\tau)$, for a range of BH mass. Our findings, thus, open up an enormous opportunity to explore the prospect of direct detection of the superradiance phenomena and a new way of observing the PBH-PBH merging phenomena.

\begin{figure}[t]
\centering
\includegraphics[scale=0.43]{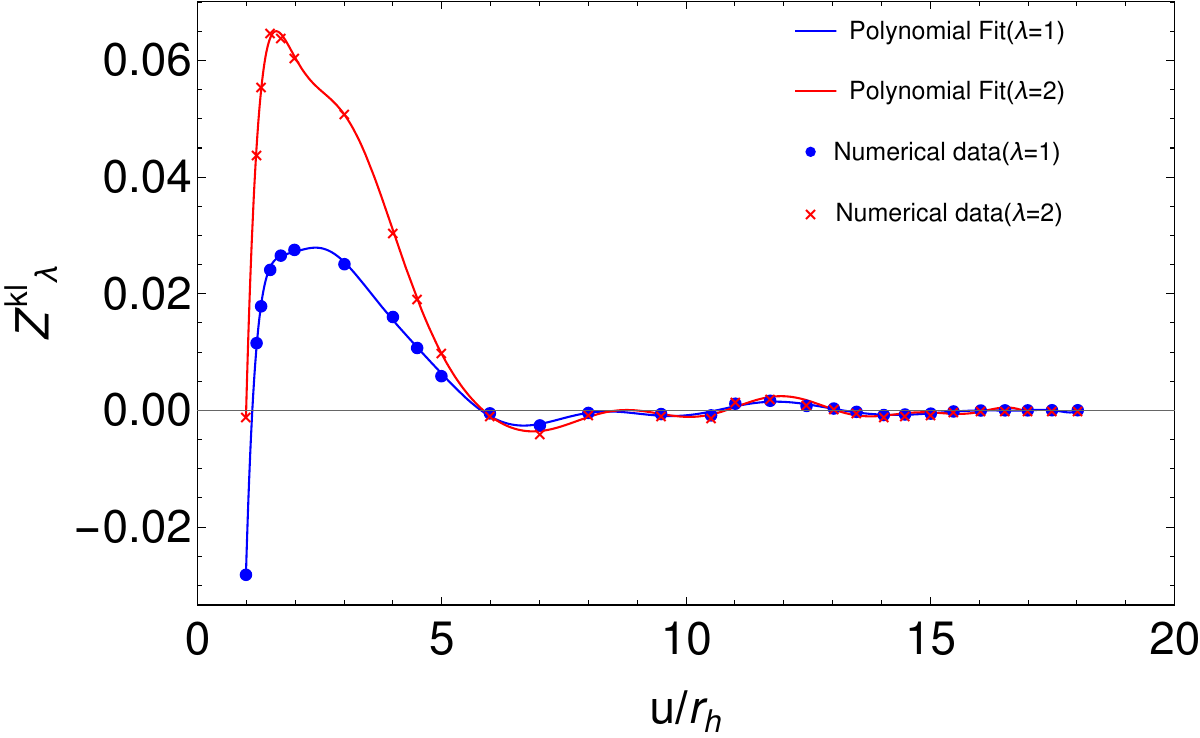}
\caption{Mean amplification factor has been plotted with time by averaging over the position of the interaction surface, $r_{\rm int}$, for a fixed frequency, $\km=0.1 r^{-1}_h$ and $l=1$. The points indicate the results of our numerical analysis, and the solid line shows the corresponding polynomial fit. The positive parts indicate the amplification, while the negative parts represent the absorption of incoming EM radiation.}\label{amp_factor}
\end{figure}

\begin{figure}[t]
\centering
\includegraphics[scale=0.43]{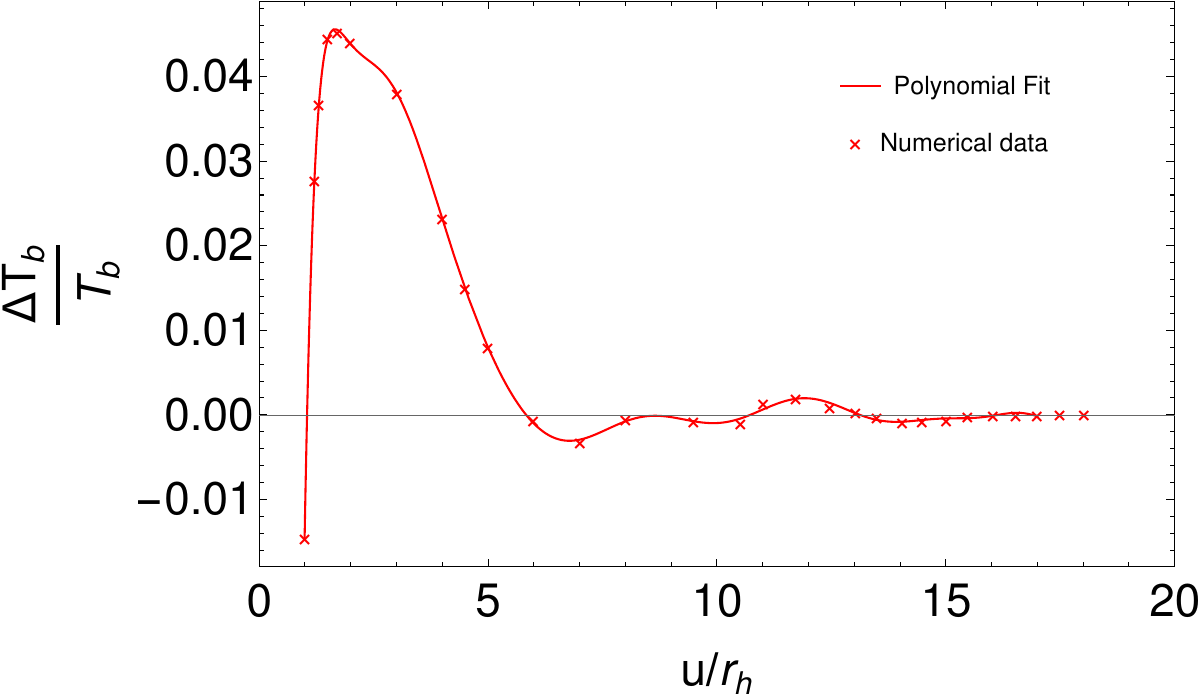}
\caption{Variation of the brightness temperature has been plotted considering the integrated effect of the amplification over the position of the interaction surface, $r_{\rm int}$, for a fixed frequency, $\km=0.1 r^{-1}_h$ and $l=1$. The points indicate the results of our numerical analysis, and the solid line shows the corresponding polynomial fit. The positive and negative values represent the increment and decrement of the brightness temperature due to the superradiant amplification and absorption, respectively.}\label{btemp}
\end{figure}

{\bf Mean amplification factor and brightness temperature:} The extracted energy by the EM field from the ringing black hole background is manifested in the mean absorption cross section through the negative part (Fig.\ref{Integrated}). Correspondingly, the amplification factor can be computed. However, the presence of mode coupling causes the symmetry between the two independent degrees of freedom to break. This is evident from the fact that the source term in the respective equation of motion, \eqref{eom_expand2}, is different. For this reason, it is natural to define the individual amplification factor for the two polarization modes as
\be\label{ampfactor1}
Z_{\lambda}^{{\km}l}(u, r_{\rm int})=\left|\frac{\mathcal{R}_\lambda(u)}{\mathcal{I}_\lambda(u)}\right|^2-1, 
\ee
where $\lambda (=1,2)$ representing the two degrees of freedom corresponding to the components $\chi_1$ and $\chi_2$ of the gauge invariant variable \eqref{invvar}. In the static black hole case, where these correspond to two polarization modes, one may refer to Ref.\cite{Crispino:2007qw}. Notably, we have omitted the $\km, l$ and $r_{\rm int}$ indices from the incident and reflected amplitudes, as done previously. Nevertheless, the dependence is still assumed. Whereas, the azimuthal mode is considered to be $m=1$, fixed from the normalization as discussed in Sec.\ref{def.ACS}. Now, the amplification factor has been defined for the individual interaction surface. For a physical realization, this should also be considered in an integrated form, as is done for the absorption cross section. We define the mean amplification factor as
\be\label{mean.amp}
Z_{\lambda}^{{\km}l}(u)=\frac{\int Z_{\lambda}^{{\km}l}(u, r_{\rm int}) dr_{\rm int}}{\int dr_{\rm int}}.
\ee
This definition for individual components is valid in the sense that they appear to be additive in \eqref{energy.extract}. To evaluate the above quantity numerically, we have followed a similar procedure to that used for the mean absorption cross section, discussed in the previous section, and plotted it in Fig.\ref{amp_factor}. The difference in amplification among the two components is distinguishable and arises because the two modes follow different dynamics due to mode–mode coupling \eqref{chieqn}.
Next, we examine how this amplification factor affects the brightness temperature of the superradiantly scattered photon. The photon’s intensity is determined by Planck’s law, 
\be
I(\km)=\frac{\km^3}{2 \pi^2} \frac{1}{e^{\frac{\km}{T_b}} - 1}.
\ee
The corresponding brightness temperature can be determined analytically using the Rayleigh–Jeans approximation in the low-frequency (radio frequency) limit and can be expressed as a function of frequency:
\be\label{temp.intens}
T_b(\km) = \frac{2 \pi^2}{\km^2} I(\km).
\ee
Previously, we discussed how the energy flux incorporates the amplification factor in \eqref{energy.extract} and \eqref{ampfactor1}. Now,  the energy flux is directly proportional to the photon intensity. Whereas the fractional change in energy flux is directly linked to the amplification factor and scales proportionally for each mode. Similar to the amplification factor for individual components, $\chi_1$ and $\chi_2$, the corresponding intensities must be treated separately, and can be quantified as $I=I_1+I_2$. Nevertheless, considering the initial intensity to be approximately equal, $I^{\rm in}_1\sim I^{\rm in}_2\sim I^{\rm in}$, the fractional change in intensity and brightness temperature can be defined as,
\be
\frac{\Delta T_b(\km, l)}{T_b(\km, l)}=\frac{\Delta I_1(\km, l)+\Delta I_2(\km, l)}{2I^{\rm in}}\simeq \frac{1}{2}\sum_{\lambda}\left(\left|\frac{\mathcal{R}_\lambda(u)}{\mathcal{I}_\lambda(u)}\right|^2-1\right)\simeq \frac{1}{2}\sum Z^{\km l}_\lambda(u). 
\ee
The last two equalities follow from the standard definition of the amplification factor\cite{Crispino:2007qw}, which can be approximately identified with the mean amplification factor given in \eqref{mean.amp}. The time-varying behaviour will thus be encoded in the brightness temperature. Regardless of the initial amplitude of the electromagnetic radiation emitted by LOFAR target sources such as quasars and blazars, if this radiation is scattered in the background of a ringing black hole, transient variations in brightness temperature will be observed. We have illustrated this temperature variation, increment and decrement in Fig.\ref{btemp}, revealing the superradiant amplification and absorption of the EM radiation.  
\section{Conclusions and future directions} 
To summarize, we have achieved superradiant scattering for the EM field during the ringing phase of the black hole. 
Unlike the static case, GW-induced superradiance phenomena are transient and the time scale is directly proportional to the GW oscillation time scale.  
We have interpreted the negative values of the absorption cross section during its fast evolution as superradiant scattering. 
\textit{Our analysis reveals that superradiance phenomena arise not only in the presence of a rotating horizon  \cite{Vicente:2018mxl}, EM wave scattered through the ringing fluctuation of the BH can also undergo superradiant enhancement.} Introducing the hypothetical interaction surface, we have been able to compute the absorption cross section, which is otherwise very difficult to define for this time-dependent BH spacetime. The physical picture we propose is that the incident wave interacts with the gravitational wave background at the position of the interaction surface, $r_{\rm int}$ and gets scattered in the static Schwarzschild spacetime. Naturally, the final absorption cross section should be averaged over the interaction surface, and we define the mean absorption cross section, which is independent of $r_{int}$. This is the precise observable quantity, which should be measured by the observer at spatial infinity. The time-dependent superradiant phenomena will be realized as repeated flashes of radiation, which will eventually diminish.

Detecting the BH superradiance itself is difficult because of its extremely weak signal. GW-induced time-varying superradiance signal, in principle, should be easier to detect than the static one. For such a case, though, the challenge arises due to its very short time scale of oscillation. We further would like to point out that BH-BH or BH-Neutron star mergers detected by LIGO, Virgo and KAGRA \cite{LIGOScientific:2016aoc, LIGOScientific:2016emj, LIGOScientific:2016vbw, LIGOScientific:2016vlm, LIGOScientific:2016sjg, LIGOScientific:2017ycc, LIGOScientific:2017vwq, LIGOScientific:2020stg, LIGOScientific:2016jlg, LIGOScientific:2020aai, LIGOScientific:2020zkf, LIGOScientific:2017vox, LIGOScientific:2023fpk, KAGRA:2023pio, KAGRA:2022twx} should have produced such transient superradiant EM signals. However, due to their mass $> M_{\odot}$ the wavelength of the signal they produced is very large of $\gtrsim 10$ km, and observing such EM waves is far from any current experimental limit. This immediately suggests us to look for the BH mass, which should be very low $M < M_{\odot}$ and consequently they can produce a detectable superradiant signal in the radio or higher frequency range. Interestingly, those mass ranges should be of primordial origin, which has gained widespread interest in the recent past as an alternative to dark matter candidates. Depending on the formation period, such BHs, known as PBHs, can have a broad mass range. PBHs with an extended mass function can still be alive today and may undergo a merging phase. As a result, these PBHs, undergoing the ringdown phase exhibit superradiance. Our analysis suggests that existing ground-based radio telescopes, such as the Low-Frequency Array (LOFAR), might be capable of detecting these transient signals from PBHs, with mass range $M\sim 10^{-1} - 10^{-2} M_{\odot}$. Furthermore, we have also estimated the brightness temperature for this transient signal, which can be used to reveal the signature of superradiant amplification. Our current results, therefore, reveal an intriguing possibility of observing PBH-PBH merger events through electromagnetic waves. 

To this end, we must reiterate the significance of our outcome, which is its complementary nature as a potential observable signature of the ringdown phase of BHs. 
Along with our previous findings for the scalar field \cite{Karmakar:2021ssg}, the effect on the EM field discussed in the present paper establishes a generic feature of the ringing BHs of having the superradiant scattering of any fundamental field. Although fermions have been shown not to exhibit superradiant enhancement \cite{Unruh:1973bda, Brito:2015oca} in case of a rotating BH, existing studies point out that certain BHs, such as extremal BH \cite{Dai:2023zcj}  or in the case of a charged AdS BH \cite{Dai:2023zcj, Dias:2019fmz}, it may be possible for the fermions to develop superradiant instability. In our future publication, we would like to take up these issues along with their direct detection prospects.

Some of the promising extensions of the present analysis that could be done are as follows. Considering higher quasinormal modes for gravitational perturbation will be interesting, and in such cases, the spacetime may differ for each mode. As previously noted, the validity of the perturbative expansion of test fields on the ringing BH spacetime must be carefully investigated. If it holds, the procedure for the overall analysis would remain essentially the same. Given the time-dependent nature, we can expect superradiant characteristics for the higher gravitational modes as well. Next, generalization to the spinning black holes, which already bear the superradiant features of the bosonic field, could be interesting to explore. Along the same line, further generalization can be done considering deformed \cite{Konoplya:2016pmh, Franzin:2021kvj, Siqueira:2022tbc} BH  spacetime and binary black hole system \cite{Wong:2019kru} in the ringing phase. Another important direction could be to consider the scattering of GW waves \cite{East:2013mfa, Rosa:2015hoa, Rosa:2016bli} itself from the ringing black holes and investigate their observable features in the current GW detectors, LIGO/VIRGO/KAGRA. We are currently working on some of the directions mentioned above. 

Studying such scattering phenomena from the time-dependent analog systems, which have characteristic similarities with the astrophysical systems\cite{Berti:2004ju, Basak:2002aw, Torres:2019sbr}, could be an interesting topic to study. There already exist some studies \cite{Torres:2016iee, Cromb:2020ldn, Jacquet:2021scv, Cardoso:2022yin, Giacomelli:2020evu} on the superradiant enhancement of fluid fluctuations in the analog background. However, in the oscillating analog system (one may look at \cite{Karmakar:2023yce}), such studies can be interesting and have not been explored in detail.\\
\noindent
{\bf Acknowledgments:} We would like to thank our
Gravity and High Energy Physics groups at IIT Guwahati for useful discussions in several contexts. RK wants to thank Md Riajul Haque for the discussions regarding PBH. Thanks to the anonymous referee for useful comments and suggestions, which have significantly improved our paper.

\newpage
\appendix
\section{Source terms in the nonhomogeneous differential equations of the EM field}\label{source}
In this section, we discuss elaborately the following set of equations, mentioned \eqref{chieqn} in the main text 
\be\label{appenchieom}
\bea
&\mathcal{L}_0\chi^{lm}_1+{\cal Q}^i_{lmc\gamma}(h)\chi^{c\gamma}_i+\bar{\cal Q}^i_{lmc\gamma}(h^*)\chi^{c\gamma}_i=0\\
&\mathcal{L}_0\chi^{lm}_2+{\cal R}^i_{lmc\gamma}(h)\chi^{c\gamma}_i+\bar{\cal R}^i_{lmc\gamma}(h^*)\chi^{c\gamma}_i=0
\eea
\ee
The expression for the source terms ${\cal Q}^i_{lmc\gamma}(h)\chi^{c\gamma}_i$ can be expressed as 
\be\label{eomQ}
{\cal Q}^i_{lmc\gamma}(h)\chi^{c\gamma}_i=\pr_r(E0)-\frac{1}{f(r)}\pr_t (E1)
\ee
where, $E_0$ and $E_1$ have the following structure: 
\be
\bea
&E0=\frac{e^{-i\omega t}}{2}\Bigg[\Lambda^{(0,0)}_{c\gamma lm}\Big\{-f(r)c(c+1)\chi^{c\gamma}_1(-2 h^{(e)}_1(2\sqrt{5}))+\frac{1}{r^2}\pr_t\chi^{c\gamma}_2 h_2(2\sqrt{5})c(c+1)\\ 
&-\chi^{c\gamma}_4Gc(c+1)2\sqrt{5}+\frac{1}{r^2}\pr_t\chi^{c\gamma}_2 h_2\frac{3}{2}\sqrt{\frac{5}{\pi}}4\gamma\frac{\pi}{3}\frac{1}{\sqrt{4\pi}}\Big\}\\
&+\Lambda^{(1,0)}_{c\gamma lm}\Bigg\{\mi \gamma\pr_t \chi^{c\gamma}_2\Big(-2(H+G)\sqrt{15}-2G\frac{3}{2}\sqrt{\frac{5}{\pi}}\Big(2\gamma\sqrt{\frac{\pi}{3}}\Big)\Big)-2f(r)\pr_t\chi^{c\gamma}_2(-\mi \gamma\pr_r h^{(e)}_1\sqrt{15})\\
&-f(r)\Big(-c(c+1)\pr_r\chi^{c\gamma}_1(2(K-3G))-2\pr_r\chi^{c\gamma}_4(-\mi \gamma h^{(o)}_1\sqrt{15})-2\chi^{c\gamma}_4(-\mi \gamma\pr_r h^{(o)}_1\sqrt{15})\\
&-2\mi \gamma H_1\pr_r \chi^{c\gamma}_2\sqrt{15}+c(c+1)\chi^{c\gamma}_12h^{(o)}_1 (-\mi \gamma\sqrt{15})+2\pr_t\pr_r \chi^{c\gamma}_2(-h^{(e)}_1\sqrt{15})\Big)\\
&-\chi^{c\gamma}_4h_2\Big(3\mi \gamma\sqrt{15}+2\mi \gamma 2\gamma\sqrt{\frac{\pi}{3}}\frac{3}{2}\sqrt{\frac{5}{\pi}}-\mi \gamma\frac{3}{2}\sqrt{\frac{5}{\pi}}2\sqrt{\frac{\pi}{3}}\Big)\Bigg\}\\
&+\Lambda^{(2,0)}_{c\gamma lm}\Bigg\{-f(r)\Big(-c(c+1)\chi^{c\gamma}_1\pr_r(2(K-3G))+c(c+1)\chi^{c\gamma}_14 h^{(e)}_1+2\chi^{c\gamma}_3 H_1c(c+1)\Big) \\
&-\chi^{c\gamma}_4\Big(-2c(c+1)H-2Gc(c+1)\Big)+\frac{1}{r^2}\pr_t\chi^{c\gamma}_2 h_2\Big(-2c(c+1)+\frac{3}{2}\sqrt{\frac{5}{\pi}}4\gamma\frac{\pi}{3}\frac{2}{\sqrt{20\pi}}\Big)\Bigg\}\\
&+\tilde{\Lambda}^{(2,0)}_{c\gamma lm}\Bigg\{-f(r)\Big(-2\pr_r\chi^{c\gamma}_4h^{(e)}_1-2\chi^{c\gamma}_4\pr_r h^{(e)}_1+c(c+1)\chi^{c\gamma}_12h^{(o)}_1-2\chi^{c\gamma}_3 H_1-4\pr_t\pr_r\chi^{c\gamma}_2h^{(o)}_1\\
&+2\pr_t\chi^{c\gamma}_2(-\pr_r h^{(o)}_1)\Big)-\chi^{c\gamma}_42(H+G)+3\frac{1}{r^2}\pr_t\chi^{c\gamma}_2 h_2\Bigg\}\\
&+\frac{1}{r^2}\pr_t\chi^{c\gamma}_2 h_2\Big\{\frac{3}{2}\sqrt{\frac{5}{\pi}}\Big(\frac{2}{3}\sqrt{2\pi}\sqrt{(c-\gamma)(c+\gamma+1)}\frac{1}{\sqrt{4\pi}}\Lambda^{(2,-1)}_{c\gamma+1 lm}\Big)\Big\}+\frac{1}{r^2}\chi^{lm}_4G2m^2\frac{3}{2}\sqrt{\frac{5}{\pi}}\Bigg]\\
\eea
\ee
and,
\be
\bea
&E1=\frac{e^{-\mi\omega t}}{2}\Bigg[\Lambda^{(0,0)}_{c\gamma lm}\Big\{-\pr_r\chi^{c\gamma}_2f(r)\frac{h_2}{r^2}\frac{3}{2}\sqrt{\frac{5}{\pi}}
4(\gamma-c(c+1))\frac{\pi}{3}\frac{1}{\sqrt{4\pi}}\\
&+2 \chi^{c\gamma}_2\Big(-c(c+1)f(r) \frac{h^{(o)}_1}{r^2}(\sqrt{5})+c(c+1)2\sqrt{\frac{\pi}{3}}f(r)\frac{h^{(o)}_1}{r^2}\sqrt{15}(\frac{1}{\sqrt{4\pi}})\Big)\\
&+f(r)G\frac{3}{2}\sqrt{\frac{5}{\pi}}\Big(-\chi^{c\gamma}_3c(c+1)4\frac{\pi}{3}\frac{1}{\sqrt{4\pi}}\Big)\Big\}+\Lambda^{(1,0)}_{c\gamma lm}\Big\{2\chi^{c\gamma}_4\gamma\omega h^{(o)}_1\sqrt{15}-\pr_r\chi^{c\gamma}_2f(r)2\mi \gamma(H-G)\sqrt{15}\\
&+2(-\mi\km)\chi^{c\gamma}_4\mi\gamma h^{(o)}_1\sqrt{15}-\pr_t\chi^{c\gamma}_2(-2 H_1\sqrt{15}\mi\gamma+2\gamma\omega h^{(e)}_1\sqrt{15})+\pr^2_t\chi^{c\gamma}_2(-2\mi\gamma h^{(e)}_1\sqrt{15})\\
&+2 \chi^{c\gamma}_2(-c(c+1)\mi\gamma f(r)h^{(e)}_1\sqrt{15})+\chi^{c\gamma}_3f(r)\mi\gamma \frac{h_2}{r^2}\Big(-2\sqrt{\frac{\pi}{3}}\frac{3}{2}\sqrt{\frac{5}{\pi}}+3\sqrt{15}-2\frac{3}{2}\sqrt{\frac{5}{\pi}}(2\gamma\sqrt{\frac{\pi}{3}})\Big)\\
&+f(r)G\frac{3}{2}\sqrt{\frac{5}{\pi}}2\mi \gamma\pr_r\chi^{c\gamma}_22\gamma\sqrt{\frac{\pi}{3}}\Big\}\\
&+\Lambda^{(1,-1)}_{c\gamma+1 lm}\Big(2\sqrt{\frac{2\pi}{3}}\sqrt{(c-\gamma)(c+\gamma+1)}\Big)\Big\{\mi\gamma\chi^{c\gamma}_3f(r)\frac{h_2}{r^2}+2\mi\gamma f(r)G\pr_r\chi^{c\gamma}_2 \frac{3}{2}\sqrt{\frac{5}{\pi}}\Big\}\\
&+\Lambda^{(2,0)}_{c\gamma lm}\Big\{2\chi^{c\gamma}_4H_1(-c(c+1))-\pr_r\chi^{c\gamma}_2f(r)\frac{h_2}{r^2}\frac{3}{2}\sqrt{\frac{5}{\pi}}
4(\gamma-c(c+1))\frac{\pi}{3}\frac{1}{\sqrt{4\pi}}\\
&+2\chi^{c\gamma}_2\Big(-c(c+1)f(r) \frac{h^{(o)}_1}{r^2}(-4)+c(c+1)2\sqrt{\frac{\pi}{3}}f(r)\frac{h^{(o)}_1}{r^2}\sqrt{15}(\frac{2}{\sqrt{20\pi}})\Big)+\chi^{c\gamma}_3f(r)\Big(2 Hc(c+1)\Big)\\
&+2(K-3G)\Big(\mi\omega c(c+1)\chi^{c\gamma}_1-c(c+1)\pr_t\chi^{c\gamma}_1\Big)+f(r)G\frac{3}{2}\sqrt{\frac{5}{\pi}}\Big(-\chi^{c\gamma}_3c(c+1)4\frac{\pi}{3}\frac{2}{\sqrt{20\pi}}\Big)\Big\}\\
&-\Lambda^{(2,-1)}_{c\gamma+1 lm}\pr_r\chi^{c\gamma}_2f(r)\frac{h_2}{r^2}\frac{3}{2}\sqrt{\frac{5}{\pi}}\Big(
\frac{2}{3}\sqrt{2\pi}\sqrt{(c-\gamma)(c+\gamma+1)}\frac{1}{\sqrt{4\pi}}\Big)\\
&+\tilde{\Lambda}^{(2,0)}_{c\gamma lm}\Big\{2\chi^{c\gamma}_4H_1+2\chi^{c\gamma}_4\mi\omega h^{(e)}_1-2\pr_t\chi^{c\gamma}_4 h^{(e)}_1-\pr_r\chi^{c\gamma}_23f(r) \frac{h_2}{r^2}-\pr_t\chi^{c\gamma}_2\{-2\mi\omega h^{(o)}_1\}+\pr^2_t\chi^{c\gamma}_2(-2h^{(o)}_1)\\
&+2 \chi^{c\gamma}_2\Big(-c(c+1)f(r)\frac{h^{(o)}_1}{r^2}\Big)
+\chi^{c\gamma}_3f(r)\Big(-2H+2G\Big)\Big\}\\
&+f(r)G\frac{3}{2}\sqrt{\frac{5}{\pi}}\chi^{lm}_3(-2m^2+l(l+1))+\pr_r \chi^{lm}_2f(r)\frac{h_2}{r^2}\frac{3}{2}\sqrt{\frac{5}{\pi}}\Big(2m^2-l(l+1)\Big)\Bigg] 
\eea
\ee
Given the expression for ${\cal Q}^i_{lmc\gamma}(h)\chi^{c\gamma}_i$, one can derive the contribution $\bar{\cal Q}^i_{lmc\gamma}(h^*)\chi^{c\gamma}_i$ coming from the complex conjugate part of the background metric by simply taking the conjugate of every metric coefficients that appeared in the previous equation for \eqref{eomQ}. As a consequence, we also have a $1/2$ factor in front of every source term.
We have made use of Wigner-3j symbols repeatedly as can be seen from the equations, with the main definition given as
\be
Y_l^m(\theta,\phi)Y_{l'}^{m'}(\theta,\phi)=\sum_{c\gamma}\Lambda^{(l',m')}_{lmc\gamma}Y_c^\gamma(\theta,\phi)
\ee
where, \[\Lambda^{(l',m')}_{lmc\gamma}=(-1)^{\gamma}\sqrt{\frac{(2l'+1)(2l+1)}{4\pi}}\sqrt{2c+1}\begin{pmatrix}
l & l' & c \\
m & m' & -\gamma\\
\end{pmatrix}\begin{pmatrix}
l & l' & c \\
0 & 0 & 0\\
\end{pmatrix}.\] For the non-zero value of the Wigner $3j$ coefficient 
$\begin{pmatrix}
l & l' & c \\
m & m' & -\gamma\\
\end{pmatrix}$,
all the $l$ values should satisfy the triangle law; the sum of any two $l$ should be greater than equal to the third one and the sum of all the $m$ values should be zero.
For the source terms in the equation \eqref{appenchieom} governing $\chi^{lm}_2$, we have, 
\begin{equation*}
\bea
&{\cal R}^i_{lmc\gamma}(h)\chi^{c\gamma}_i=\frac{e^{-i\omega t }}{2}\Bigg[\Lambda^{(0,0)}_{c\gamma lm}\Big\{\chi^{c\gamma}_3\Big(-\frac{2f(r)h_2}{r^3}+\frac{(h_2 f'(r)+f(r)\pr_r h_2)}{r^2}\Big)2\sqrt{5}\gamma(\gamma+1)\\
&+\chi^{c\gamma}_3\frac{2 f(r) h^{(o)}_1 }{r^2}2\gamma(\gamma+1)\sqrt{5}+\pr_r \chi^{c\gamma}_3\frac{f(r) h_2 }{r^2}2\sqrt{5}\gamma(\gamma+1)+\chi^{lm}_4\frac{\mi\omega}{r^2f(r)}h_22\sqrt{5}\gamma(\gamma+1)\\
&+\mi\km\chi^{c\gamma}_4\frac{1}{r^2f(r)}h_22\sqrt{5}\gamma(\gamma+1)-\Big(2 i \omega\frac{\gamma(\gamma+1)}{r^2}\chi^{c\gamma}_1(t,r)-\frac{\gamma(\gamma+1)}{r^2}\pr_t\chi^{c\gamma}_1(t,r)\Big)h^{(o)}_1(\sqrt{10}-\sqrt{5})\\
&-\chi^{c\gamma}_2(t,r)2\gamma(\gamma+1)\Big(\frac{2}{r^3}f(r)h^{(e)}_1-\frac{1}{r^2}\pr_r h^{(e)}_1-\frac{1}{r^2}h^{(e)}_1 f'(r)\Big)(\sqrt{10}-\sqrt{5})\\
&+\chi^{c\gamma}_2(t,r)\gamma(\gamma+1)\frac{1}{r^2}2(K-3G)(\sqrt{10}-\sqrt{5})+\pr_r \chi^{lm}_2(t,r)\Big(f'(r)l(l+1)2G\sqrt{5}\\
&-f(r)\frac{1}{r^2}2h^{(e)}_12l(l+1)\sqrt{5}+f(r)\frac{1}{r^2}2l(l+1)h^{(e)}_1(\sqrt{10}-\sqrt{5})-4 i \omega H_1l(l+1)\sqrt{5}\Big)\\
&+\pr^2_r \chi^{lm}_2(t,r)f(r)l(l+1)2G\sqrt{5}\Big\}+\Lambda^{(1,0)}_{c\gamma lm}\Big\{2\gamma\omega H_1\sqrt{15}\chi^{c\gamma}_3(t,r)-6f(r)\mi\gamma\sqrt{\frac{5}{3}}\pr_rG\chi^{c\gamma}_3(t,r)\\
&-f'(r)2G\sqrt{15}\mi\gamma\chi^{c\gamma}_3(t,r)-2(H f'(r)+f(r)\pr_r H)\mi\gamma\sqrt{15}\chi^{c\gamma}_3(t,r)\\
&+\frac{2 f(r) h^{(e)}_1}{r^2}(2+l(l+1))\sqrt{15}\mi\gamma \chi^{c\gamma}_3(t,r)\\
&-f(r)2G\sqrt{15}\mi\gamma\pr_r\chi^{c\gamma}_3(t,r)-f(r)2\sqrt{15} H\mi \gamma\pr_r\chi^{c\gamma}_3(t,r)-\chi^{c\gamma}_4(t,r)2\mi \gamma\sqrt{15}\pr_r H_1\\
&-2\chi^{c\gamma}_4(t,r)\sqrt{15}\frac{\gamma\omega}{f(r)}(H-G)+\sqrt{15}2\mi m H_1\pr_r\chi^{lm}_4(t,r)+2G\sqrt{15}\mi\gamma\pr_t \chi^{c\gamma}_4(t,r)\\
&+\Big(2i\omega\frac{\gamma(\gamma+1)}{r^2}\chi^{c\gamma}_1(t,r)-\frac{\gamma(\gamma+1)}{r^2}\pr_t\chi^{c\gamma}_1(t,r)\Big)\sqrt{15}\mi\gamma h^{(e)}_1\\
&-\Big(\frac{2f(r)h^{(e)}_1}{r^3}-\frac{(h^{(e)}_1 f'(r)+f(r)\pr_r h^{(e)}_1)}{r^2}\Big)\sqrt{15}\mi\gamma\chi^{c\gamma}_2(t,r)\\
&+\pr_r \chi^{c\gamma}_2(t,r)\Big(-\frac{2f(r)h_2}{r^3}+\frac{(h_2 f'(r)+f(r)\pr_r h_2)}{r^2}\Big)2\sqrt{15}\mi\gamma\\
&+2\pr_r \chi^{c\gamma}_2(t,r)f(r)\frac{1}{r^2}h^{(o)}_1\Big((2+l(l+1))\sqrt{15}\mi\gamma+\sqrt{15}l(l+1)\mi\gamma\Big)\\
&+\pr^2_r \chi^{c\gamma}_2(t,r)f(r)\frac{1}{r^2}h_22\sqrt{15}\mi\gamma+\pr_t\chi^{c\gamma}_2(t,r)\mi\omega \frac{h_2}{r^2f(r)}2\sqrt{15}\mi\gamma-\pr^2_t\chi^{c\gamma}_2(t,r)\frac{1}{r^2f(r)}h_22\sqrt{15}\mi\gamma\Big\}\\
&+\Lambda^{(2,0)}_{c\gamma lm}\Big\{-2\chi^{c\gamma}_3(t,r)\Big(-\frac{2f(r)h_2}{r^3}+\frac{(h_2 f'(r)+f(r)\pr_r h_2)}{r^2}\Big)\gamma(\gamma+1)-8\chi^{c\gamma}_3(t,r)\frac{2 f(r) h^{(o)}_1 }{r^2}l(l+1)\\
&-2\pr_r\chi^{c\gamma}_3(t,r)\frac{f(r) h_2 }{r^2}\gamma(\gamma+1)-\chi^{\gamma}_4(t,r)\frac{\mi\omega}{r^2f(r)}h_22\gamma(\gamma+1)\\
&+\pr_t \chi^{c\gamma}_4(t,r)\Big(-\sqrt{15}\frac{\mi \gamma}{f(r)}2Hr^2+\frac{1}{r^2f(r)}h_22\gamma(\gamma+1)\Big)\\
&-\Big(2 i \omega\frac{\gamma(\gamma+1)}{r^2}\chi^{c\gamma}_1(t,r)-\frac{\gamma(\gamma+1)}{r^2}\pr_t\chi^{c\gamma}_1(t,r)\Big)h^{(o)}_1(2\sqrt{2}+4)\\
&+\chi^{c\gamma}_2(t,r)\Big(-2\gamma(\gamma+1)\Big(\frac{2}{r^3}f(r)h^{(e)}_1-\frac{1}{r^2}\pr_r h^{(e)}_1-\frac{1}{r^2}h^{(e)}_1 f'(r)\Big)(2\sqrt{2}+4)\\
\eea
\end{equation*}
\be
\bea
&+l(l+1)\frac{1}{r^2}2(K-3G)(2\sqrt{2}+4)\Big)+\pr_r\chi^{c\gamma}_2(t,r)\Big(-\gamma(\gamma+1)f'(r)H(r)+\gamma(\gamma+1)2G\\
&+f(r)\frac{1}{r^2}2h^{(e)}_18\gamma(\gamma+1)+f(r)\frac{1}{r^2}2\gamma(\gamma+1)h^{(e)}_1(2\sqrt{2}+4)+2i\omega H_18\gamma(\gamma+1)\Big)\\
&-\pr^2_r \chi^{c\gamma}_2(t,r)f(r)\gamma(\gamma+1)2G-\pr_t \chi^{c\gamma}_2(t,r)\Big(-2\pr_r H_1\gamma(\gamma+1)-\mi\omega\frac{1}{f(r)}2\gamma(\gamma+1)(H-G)\Big)\\
&+\pr^2_t \chi^{c\gamma}_2(t,r)\frac{1}{f(r)}\gamma(\gamma+1)2(G-H)+4\pr_t\pr_r \chi^{c\gamma}_2(t,r) H_1\gamma(\gamma+1)\Big\}\\
&+\tilde{\Lambda}^{(2,0)}_{c\gamma lm}\Big\{2\chi^{c\gamma}_3(t,r)\Big(-\frac{2f(r)h_2}{r^3}+\frac{(h_2 f'(r)+f(r)\pr_r h_2)}{r^2}\Big)-\chi^{c\gamma}_3(t,r)\frac{4f(r) h^{(o)}_1 }{r^2}\gamma(\gamma+1)(3+\gamma(\gamma+1))\\
&+2\pr_r\chi^{c\gamma}_3(t,r)\frac{f(r) h_2 }{r^2}+2\chi^{c\gamma}_4(t,r)\frac{\mi\omega}{r^2f(r)}h_2-2\pr_t \chi^{c\gamma}_4(t,r)\frac{1}{r^2f(r)}h_2\\
&+\Big(2 i \omega\frac{\gamma(\gamma+1)}{r^2}\chi^{c\gamma}_1(t,r)-\frac{\gamma(\gamma+1)}{r^2}\pr_t\chi^{c\gamma}_1(t,r)\Big)h^{(o)}_1\\
&+\chi^{c\gamma}_2(t,r)2\gamma(\gamma+1)\Big(\frac{2}{r^3}f(r)h^{(e)}_1-\frac{1}{r^2}\pr_r h^{(e)}_1-\frac{1}{r^2}h^{(e)}_1 f'(r)\Big)+\pr_r\chi^{c\gamma}_2(t,r)\Big(f'(r)H(r)+f'(r)2G\\
&-f(r)\frac{1}{r^2}2h^{(e)}_1(3+\gamma(\gamma+1))-f(r)\frac{1}{r^2}2l(l+1)h^{(e)}_1-f(r)2\pr_r(K-3G)-2i\omega H_1(3+\gamma(\gamma+1))\Big)\\
&+\pr^2_r \chi^{c\gamma}_2(t,r)\Big(f(r)2G+2f(r)H\Big)-\pr_t \chi^{c\gamma}_2(t,r)\Big(2\pr_r H_1+\mi\omega\frac{1}{f(r)}2(H-G)\Big)\\
&+\pr^2_t \chi^{c\gamma}_2(t,r)\frac{1}{f(r)}2(H-G)-4\pr_t\pr_r \chi^{c\gamma}_2(t,r) H_1\Big\}\\
&+S_{c\gamma lm}\Big\{\chi^{c\gamma}_3(t,r)\Big(-f'(r)3\mi\gamma G\sqrt{\frac{5}{\pi}}+\frac{2 f(r) h^{(e)}_1}{r^2}3\sqrt{\frac{5}{\pi}}\mi \gamma \Big)-\pr_r\chi^{c\gamma}_3(t,r)f(r)\Big(3\mi \gamma G\sqrt{\frac{5}{\pi}}\Big)\\
&+\pr_t \chi^{c\gamma}_4(t,r)\Big(3\mi \gamma G\sqrt{\frac{5}{\pi}}\Big)r^2+\pr_r \chi^{c\gamma}_2(t,r)\Big(\Big(-\frac{2f(r)h_2}{r^3}+\frac{(h_2 f'(r)+f(r)\pr_r h_2)}{r^2}\Big)\Big(3\mi \gamma\sqrt{\frac{5}{\pi}}\Big)\\
&+2f(r)\frac{1}{r^2}h^{(o)}_1\Big(3\sqrt{\frac{5}{\pi}}\mi\gamma\Big)\Big)-\pr^2_r \chi^{c\gamma}_2(t,r)f(r)\frac{1}{r^2}h_2\Big(-3\mi \gamma\sqrt{\frac{5}{\pi}}\Big)-\pr_t \chi^{c\gamma}_2(t,r)\mi\omega \frac{h_2}{r^2f(r)}\Big(-3\mi\gamma\sqrt{\frac{5}{\pi}}\Big)\\
&+\pr^2_t \chi^{c\gamma}_2(t,r) \frac{1}{r^2f(r)}h_2\Big(-3\mi\gamma\sqrt{\frac{5}{\pi}}\Big)\Big\}\\
&+\tilde{S}_{c\gamma lm}\Big\{-\chi^{c\gamma}_3(t,r)\frac{2 f(r) h^{(o)}_1 }{r^2}2\gamma(\gamma+1)\frac{3}{2}\sqrt{\frac{5}{\pi}}-\pr_r \chi^{c\gamma}_2(t,r)\Big(f(r)\frac{1}{r^2}2h^{(e)}_1\frac{3}{2}\sqrt{\frac{5}{\pi}}+2 i \omega H_1\frac{3}{2}\sqrt{\frac{5}{\pi}}\Big)\Big\}\\
&+\chi^{lm}_3(t,r)\Big\{\Big(-\frac{2f(r)h_2}{r^3}+\frac{(h_2 f'(r)+f(r)\pr_r h_2)}{r^2}\Big)\Big(-3m^2\sqrt{\frac{5}{\pi}}\Big)+\frac{2 f(r) h^{(o)}_1 }{r^2}2l(l+1)3m^2\sqrt{\frac{5}{\pi}}\Big\}\\
&+\pr_r\chi^{lm}_3(t,r)\Big\{\frac{f(r) h_2 }{r^2}\Big(-3m^2\sqrt{\frac{5}{\pi}}\Big)\Big\}+\chi^{lm}_4(t,r)\Big\{\frac{\mi\omega}{r^2f(r)}h_2\Big(-3m^2\sqrt{\frac{5}{\pi}}\Big)\Big\}\\
&+\pr_t \chi^{lm}_4(t,r)\Big\{-\frac{1}{r^2f(r)}h_2\Big(-3m^2\sqrt{\frac{5}{\pi}}\Big) \Big\}+\chi^{lm}_2(t,r)\Big\{-l^2(l+1)^22(K-3G)\Big)\Big)\Big\}\\
&+\pr_r \chi^{lm}_2(t,r)\Big\{+2(f(r) H'(r)-f(r)\Big(-2l(l+1)\pr_r(K_3G)\Big)\\
&-f'(r)3m^2G\sqrt{\frac{5}{\pi}}+f(r)\frac{1}{r^2}2h^{(e)}_1\Big(3m^2\sqrt{\frac{5}{\pi}}\Big)+2 i \omega H_1\Big(+3m^2\sqrt{\frac{5}{\pi}}\Big)\Big\}\\
&+\pr^2_r \chi^{lm}_2(t,r)\Big\{-f(r)\Big(3m^2G\sqrt{\frac{5}{\pi}}\Big)\Big\}+\pr^2_t \chi^{lm}_2(t,r)\Big\{\frac{1}{f(r)}\Big(3m^2G\sqrt{\frac{5}{\pi}}\Big)\Big\}
\Bigg]
\eea
\ee
Where we have used the following mathematical quantities, constructed out of $\Lambda^{(l',m')}_{lmc\gamma}$, are 
\be
\bea
&S_{lmc\gamma}\equiv 2m\sqrt{\frac{\pi}{3}}\sum_{c\gamma}\Lambda^{(1,0)}_{lmc\gamma}+2\sqrt{\frac{2\pi}{3}}\sqrt{(l-m)(l+m+1)}\Lambda^{(1,-1)}_{lm+1c\gamma}\\
&\tilde{S}_{lmc\gamma}\equiv\frac{m}{3}\sqrt{4\pi}\Lambda^{(0,0)}_{lmc\gamma}+\frac{2m}{3}\sqrt{\frac{4\pi}{5}}\Lambda^{(2,0)}_{lmc\gamma}+4\sqrt{2}\frac{\pi}{3}\sqrt{(l-m)(l+m+1)}\frac{1}{\sqrt{4\pi}}\Lambda^{(2,-1)}_{lm+1c\gamma}\\
&\tilde{\Lambda}^{(2,0)}_{lmc\gamma}\equiv-\Big[2m\Lambda^{(2,0)}_{lmc\gamma}+m\sqrt{5}\Lambda^{(0,0)}_{lmc\gamma}+\sqrt{(l-m)(l+m+1)}3\sqrt{\frac{2}{3}}\Lambda^{(2,-1)}_{lm+1c\gamma}\Big]
\eea
\ee
Note that, given the expression for ${\cal R}^i_{lmc\gamma}(h)\chi^{c\gamma}_i$ one can derive the contribution $\bar{\cal R}^i_{lmc\gamma}(h)\chi^{c\gamma}_i$ in the same way described for $\bar{\cal Q}^i_{lmc\gamma}(h)\chi^{c\gamma}_i$.
\section{Evaluation of the normalization factors}\label{normfac}
In the following discussion, our main focus will be to calculate the incident energy density and to formulate the procedure to calculate the normalization factors.  We start with a circularly polarized \cite{Jackson:1998nia} ingoing plane EM wave propagating along $z$-direction towards the black hole as,
\be
\bea
&A_x(u,{\bf x})=e^{-\mi\km(u+2z)}\\
&A_y(u,{\bf x})=\mi e^{-\mi\km(u+2z)} .
\eea
\ee
The approach to finding out the normalization factor is to compare the incident plane wave with the asymptotic form of the field solution, which is obtained in spherical coordinates. We, therefore, need to transform the plane EM wave written in cartesian coordinate, $\textbf{A}=A_{x}\hat{x}+A_{y}\hat{y}$, to spherical coordinate
by using, $A'_{\mu}(x')=(\partial x^{\nu}/\partial x'^{\mu})A_{\nu}(x)$ as, 
\be\label{pln.sph1}
\bea
&A'_u(u,\textbf{r})=A_u(u,\textbf{x}) ;~~~A'_r(u,\textbf{r})=\sin\theta e^{\mi\phi}A_x(u,{\xb})\\
&A'_\theta(u,\textbf{r})=r\cos\theta e^{\mi\phi}A_x(u,{\xb}) ;~~~A'_\phi(u,\textbf{r})=ir\sin\theta e^{\mi\phi}A_x(u,{\xb}),\\
\eea
\ee
where, we have used $A_{z}=0$ and $A_y=\mi A_x$. Of course, $A_x(u,\xb)$ should be expressed in spherical coordinates using Rayleigh expansion of the spatial part of a plane wave propagating along the z-direction given by
\be
e^{-\mi\km z}=e^{-\mi\km r\cos\theta}=\sum_{l=0}(2l+1)\mi^l j_{l}(\km r)P^0_l(\cos\theta) .
\ee
Taking the derivative with respect to $\theta$ on both sides of the above equation   one can discover the following expression also 
\be
e^{-\mi\km z}=e^{-\mi\km r\cos\theta}=\sum_{l=0}(2l+1)\mi^l \frac{j_{l}(\km r)}{\mi\km r}\frac{\pr_\theta P^0_l(\cos\theta)}{\sin\theta}.
\ee
Where one needs to use the relation $P^1_l(\cos\theta)=-\pr_\theta P^0_l(\cos\theta)$. Finally, one obtains the components of plane EM waves for circularly polarized light in spherical coordinates as,
\be\label{pln.sph2}
\bea
&A'_u(u,\textbf{r})=\sum_{lm} A'^{lm}_u(u,r) Y_{lm}(\Omega)=0\\
&A'_r(u,\textbf{r})=\sum_{lm} A'^{lm}_r(u,r)Y_{lm}(\Omega)=\sum_{lm}(-1)^{l+1}\delta_{m1}\sqrt{4\pi(2l+1)l(l+1)}\frac{e^{-\mi\km(u+2r_*)}}{2\km^2r^2}Y_{lm}(\Omega)\\
&~~~~~~~~~~~~~~~~~~~~~~~~~~~~~~~~~~~~~~~~~~~~~~~+outgoing~part\\
&A'_{s}(u,{\rb})=-\sum_{lm}(-1)^{l+1}\delta_{m1}\sqrt{\frac{4\pi(2l+1)}{l(l+1)}}\Big[\mi\Psi^{lm}_{s}(\Omega)+\Phi^{lm}_{s}(\Omega)\Big]\frac{e^{-\mi\km(u+2r_*)}}{2k}+outgoing~part\\
 \eea
 \ee
 Note that the $\delta_{m1}$ factor in the above expressions comes due to $e^{i \phi}$ in the $A'_u$ and $A'_{\phi}$ components of the gauge field (see eqs.\eqref{pln.sph1} and \eqref{pln.sph2}). The above expansions could be checked following $j_{l}(\km r)+\mi n_{l}(\km r)\sim\frac{(-\mi)^{l+1}e^{\mi\km r}}{\km{r}},~j_{l}(\km r)-\mi n_{l}(\km r)\sim\frac{\mi^{l+1} e^{-\mi\km r}}{\km{r}}$ for $r\to\infty$ or $\km r>>1$, that implies $j_l(\km r)\sim i^{l+1}\frac{e^{-\mi\km r}}{2\km r}+(-i)^{l+1}\frac{e^{\mi\km r}}{2\km r}$.
Gauge invariant variables for circularly polarized light in this new gauge become
\be\label{inwave}
\bea
&\sum_{lm}\tilde{\chi}^{\km lm}_1(u,r)Y^{lm}(\Omega)=\sum_{lm}\frac{r^2}{l(l+1)}(\pr_u A'^{lm}_r-\pr_r A'^{lm}_u)\\
&~~~~~~~~~~~~~~~~~~~~~~~~~~~~~~~~~~~~~~~~=-\mi\sum_{lm}(-1)^{l+1}\delta_{m1}\sqrt{\frac{4\pi(2l+1)}{l(l+1)}}\frac{e^{-\mi\km(u+2r_*)}}{2\km}Y_{lm}(\Omega)+out~going\\
\eea
\ee
Also in the expansion of $A'_{s}(u,{\rb})$ \eqref{pln.sph2} one can check that the term with $\Phi^{lm}_{s}(\Omega)$ does not transform, and we denote this as,
\be
\bea
&\sum_{lm}\tilde{\chi}^{\km lm}_2(u,r)\Phi^{lm}_s(\Omega)=-\sum_{lm}(-1)^{l+1}\delta_{m1}\sqrt{\frac{4\pi(2l+1)}{l(l+1)}}\frac{e^{-\mi\km(u+2r_*)}}{2\km}\Phi^{lm}_{s}(\Omega)+out~going\\
\eea
\ee
(recall the $\frac{r^2}{l(l+1)}$ factor in the definition of $\tilde{\chi}_1$ from the main text). 
With the above gauge invariant variables, one can deduce the normalization factors from the asymptotic form of these fields,
\be\label{asympformchi1chi2}
\bea
&\sum_{lm}\tilde{\chi}^{\km lm}_1(u,r)Y^{lm}(\Omega)=\sum_{lm}\mathcal{N}^{\km lm}_1[\mathcal{I}_1(u)e^{-\mi\km(u+2r_*)}+\mathcal{R}_1(u)e^{-\mi\km u}]Y_{lm}(\Omega), \\
&\sum_{lm}\tilde{\chi}^{\km lm}_2(u,r)\Phi^{lm}_s(\Omega)=\sum_{lm}\mathcal{N}^{\km lm}_2[\mathcal{I}_2(u)e^{-\mi\km(u+2r_*)}+\mathcal{R}_2(u)e^{-\mi\km u}]\Phi^{lm}_{s}(\Omega).
\eea
\ee
Comparing with \eqref{inwave} we fix the normalization factor to be
\be
\bea
&\mathcal{N}^{\km lm}_1=-\mi(-1)^{l+1}\delta_{m1}\sqrt{\frac{4\pi(2l+1)}{l(l+1)}}\frac{1}{2\km \mathcal{I}_1(u\to \infty)},\\
&\mathcal{N}^{\km lm}_2=(-1)^l\delta_{m1}\sqrt{\frac{4\pi(2l+1)}{l(l+1)}}\frac{1}{2\km\mathcal{I}_2(u\to \infty)}.
\eea
\ee
We proceed with this crude approximation, considering only the effect in the static limit $u\to \infty$ and time dependent effect in the energy density will come from the rest of the part of the solution. We follow the definition of the incident energy density discussed in the previous section,
\be\label{fluxzu.in}
\pr_u\mathcal{G}_{in}={\mathcal{T}^z}_u=[{\mathcal{T}_{uz}}-{\mathcal{T}_{uu}}],
\ee
but we will transform this quantity to the spherical coordinate using,
\be
\mathcal{T}'_{\mu\nu}(x')=\left(\partial x^{\alpha}/\partial x'^{\mu}\right)\left(\partial x^{\beta}/\partial x'^{\nu}\right)\mathcal{T}_{\alpha\beta}(x)
\ee
So, after simplification,
\be\label{fluxsu.in}
\bea
\pr_u\mathcal{G}=\Big\{\frac{1}{2}g^{ss}(\pr_u A_s-\pr_s A_u)(\pr_r A^*_s-\pr_s A^*_r)+c.c.\Big\}-g^{ss}(\pr_u A_s-\pr_s A_u)(\pr_u A^*_s-\pr_s A^*_u)
\eea
\ee
and as the incident plane wave is considered to be propagating along the z-direction we suitably choose, $\theta\to 0$, as per the requirement.

In the following discussion, we want to express the gauge invariant combinations taking only the ingoing part of \eqref{asympformchi1chi2},
\be
\bea
&\pr_u A_s-\pr_s A_u=\sum_{lm}\left[\left(\pr_r\tilde{\chi}^{lm}_1-\pr_u\tilde{\chi}^{lm}_1\right)\Psi^{lm}_s(\Omega)+\pr_u\tilde{\chi}^{lm}_2\Phi^{lm}_s(\Omega)\right]\\
&=\mi\sum_{lm}(-1)^{l+1}\delta_{m1}\sqrt{\frac{4\pi(2l+1)}{l(l+1)}}\Big[\frac{\mi}{\km}\frac{d}{du}\left(\frac{\mathcal{I}^{lm}_2(u)}{\mathcal{I}^{lm}_2(u\to\infty)}\right)\Phi^{lm}_{s}+\frac{\mathcal{I}^{lm}_2(u)}{\mathcal{I}^{lm}_2(u\to\infty)}\Phi^{lm}_{s}\\
&~~~~~~~~~~~~~~~~~~~~~~+\frac{\mi}{\km}\frac{d}{du}\left(\frac{\mathcal{I}^{lm}_1(u)}{\mathcal{I}^{lm}_1(u\to\infty)}\right)\Psi^{lm}_{s}+\frac{\mathcal{I}^{lm}_1(u)}{\mathcal{I}^{lm}_1(u\to\infty)}\mi\Psi^{lm}_{s}\Big]\frac{e^{-\mi\km(u+2r)}}{2}+outgoing~part\\
&\pr_r A_s-\pr_s A_r=\sum_{lm}\left[\pr_r\tilde{\chi}^{lm}_1\Psi_s+\pr_r\tilde{\chi}^{lm}_2\Phi_s\right]\\
&=2\mi\sum_{lm}(-1)^{l+1}\delta_{m1}\sqrt{\frac{4\pi(2l+1)}{l(l+1)}}\left[\frac{\mathcal{I}^{lm}_2(u)}{\mathcal{I}^{lm}_2(u\to\infty)}\Phi^{lm}_{s}(\Omega)+\frac{\mathcal{I}^{lm}_1(u)}{\mathcal{I}^{lm}_1(u\to\infty)}\mi\Psi_s^{lm}(\Omega)\right]\frac{e^{-\mi\km(u+2r)}}{2}\\
&~~~~~~~~+outgoing~part\\
\eea
\ee
where we have used the equations governing the gauge invariant variables in $(u,r)$ coordinate for asymptotic Schwarzschild space-time (One may look at Appendix.\ref{eomucoord}) as our aim to calculate the absorption cross section at $r\to \infty$. We express the above invariant combinations in Cartesian Coordinates in the following way :
\be\label{sph.cart}
\bea
&\pr_u A_s-\pr_s A_u\\
&=\frac{1}{\sum_{lm}(-1)^{l+1}\delta_{m1}\sqrt{\frac{4\pi(2l+1)}{l(l+1)}}[\Phi^{lm}_s(\Omega)+\mi\Psi^{lm}_s(\Omega)]}\sum_{lm}(-1)^{l+1}\delta_{m1}\sqrt{\frac{4\pi(2l+1)}{l(l+1)}}\times\\
&\left[\frac{\mi}{\km}\frac{d}{du}\left(\frac{\mathcal{I}^{lm}_2(u)}{\mathcal{I}^{lm}_2(u\to\infty)}\right)\Phi^{lm}_{s}+\frac{\mathcal{I}^{lm}_2(u)}{\mathcal{I}^{lm}_2(u\to\infty)}\Phi^{lm}_{s}+\frac{\mi}{\km}\frac{d}{du}\left(\frac{\mathcal{I}^{lm}_1(u)}{\mathcal{I}^{lm}_1(u\to\infty)}\right)\Psi^{lm}_{s}+\frac{\mathcal{I}^{lm}_1(u)}{\mathcal{I}^{lm}_1(u\to\infty)}\mi\Psi^{lm}_{s}\right]\times\\
&\times(-\mi\km A'_s(u,\textbf{r}))\\
&=\xi_{us}(k,u)(-\mi\km A'_s(u,\textbf{r}))\\
&\pr_r A_s-\pr_s A_r=\frac{\sum_{lm}(-1)^{l+1}\delta_{m1}\sqrt{\frac{4\pi(2l+1)}{l(l+1)}}\left[\frac{\mathcal{I}^{lm}_2(u)}{\mathcal{I}^{lm}_2(u\to\infty)}\Phi^{lm}_{s}(\Omega)+\frac{\mathcal{I}^{lm}_1(u)}{\mathcal{I}^{lm}_1(u\to\infty)}\mi\Psi_s^{lm}(\Omega)\right]}{{\sum_{lm}(-1)^{l+1}\delta_{m1}\sqrt{\frac{4\pi(2l+1)}{l(l+1)}}[\Phi^{lm}_s(\Omega)+\mi\Psi^{lm}_s(\Omega)]}}(-2\mi\km A'_s(u,\textbf{r}))\\
&=\xi_{rs}(k,u)(-2\mi\km A'_s(u,\textbf{r}))
\eea
\ee
where, $\xi_{rs}(k,u), \xi_{rs}(k,u)$ has to be evaluated numerically, and $A'_s(u,\textbf{r}))$ is the spherical representation of the plane wave in flat space. Substituting the above combinations in the expression of incident energy density we arrive at,   
\be
\bea
\pr_u\mathcal{G}&=\Big\{\frac{1}{2}g^{ss}(\pr_u A_s-\pr_s A_u)(\pr_r A^*_s-\pr_s A^*_r)+c.c.\Big\}-g^{ss}(\pr_u A_s-\pr_s A_u)(\pr_u A^*_s-\pr_s A^*_u)\\
&=2\Big\{\frac{\km^2}{2}g^{ss}\xi_{us}(k,u)\xi^*_{rs}(k,u)A'_s(u,\textbf{r}){A^*}'_s(u,\textbf{r})+c.c.\Big\}-\km^2g^{ss}|\xi_{us}(k,u)|^2A'_s(u,\textbf{r}){A^*}'_s(u,\textbf{r})\\
\eea
\ee
As a consistency check, one may look for the case of static black hole limit, which corresponds to $u\to\infty$. In this limit the components \eqref{sph.cart} become
\be
\bea
&\pr_u A_s-\pr_s A_u=-\mi\km A'_s(u,\textbf{r})\\
&\pr_r A_s-\pr_s A_r=-2\mi\km A'_s(u,\textbf{r})\\
\eea
\ee
by substituting these combinations
\be
\bea
\pr_u\mathcal{G}&=2\km^2\cos^2\theta+2\km^2-\km^2\cos^2\theta-\km^2=2\km^2
\eea
\ee
and as the incident plane wave is considered to be propagating along the z-direction we suitably choose, $\theta\to 0$.
\section{Gauge field equations for static Schwarzschild, ($t$,$r$)-coordinate}\label{schw.eom.tr}
Substituting the field components \eqref{modedecom} in the equation of motion \eqref{Eom} of the gauge field, for static Schwarzschild space time dictated by the metric $g^{\mu\nu}_0(t,r)$ we obtain the following set of equation,
\be
\bea
&f(r)\frac{1}{l(l+1)}\pr_r(r^2(\pr_t d^{lm}-\pr_r b^{lm}))+(b^{lm}-\pr_t k^{lm})=0,\\
&\frac{r^2}{l(l+1)}\pr_t(\pr_t d^{lm}-\pr_r b^{lm})+f(r)(d^{lm}-\pr_r k^{lm})=0,\\
& f(r)\pr_r(f(r)\pr_r a^{lm})-\pr^2_t a^{lm}-f(r)\frac{l(l+1)}{r^2}a^{lm}=0,
\eea
\ee
which can be simplified and expressed in terms of the gauge invariant variables \eqref{invvar} as,
\be\label{apnd.constr}
\bea
& f(r)\pr_r\chi^{lm}_1+\chi^{lm}_4=0\\
&\pr_t \chi^{lm}_1+f(r)\chi^{lm}_3=0\\
& f(r)\pr_r(f(r)\pr_r \chi^{lm}_2)-\pr^2_t \chi^{lm}_2-f(r)\frac{l(l+1)}{r^2}\chi^{lm}_2=0
\eea
\ee
The first two of the above set of equations can be decoupled to obtain the governing equation of $\chi^{lm}_1$ as,
\be
f(r)\pr_r(f(r)\pr_r\chi^{lm}_1)-\pr^2_t\chi^{lm}_1-f(r)\frac{l(l+1)}{r^2}\chi^{lm}_1=0
\ee
\section{Gauge Field Equation in $(u,r)$ Coordinate For Schwarzschild BH}\label{eomucoord}
In this section, we discuss the coordinate transformation from $(t,{\bf r})$ to $(u(=t-r_*),{\bf r})$ and derive the expressions for EM field components and their invariant combinations. Consider the general definition of coordinate transformation for rank-1 tensor, from $x^\mu(t,r)\to x'^\mu(u,r)$, given as \cite{Padmanabhan:2010zzb}, 
\be
\tilde{A}_{\mu}(x')=\frac{\partial x^{\nu}}{\partial x'^{\mu}}A_{\nu}(x)
\ee
under which the components \eqref{modedecom} of the EM wave transform as
\be
\bea
&\tilde{A}_u(u,{\bf r})=\frac{\partial t}{\partial u}A_t(t,{\bf r})=A_t(t,{\bf r})=b^{lm}(t,r)Y_{lm}(\Omega)\\
&\tilde{A}_r(u,\textbf{r})=\frac{\partial t}{\partial r}A_t(t,\textbf{r})+\frac{\partial r}{\partial r}A_r(t,\textbf{r})=\frac{1}{f(r)}A_t(t,\textbf{r})+A_r(t,\textbf{r})=\Big(d^{lm}(t,r)+\frac{b^{lm}(t,r)}{f(r)}\Big)Y_{lm}(\Omega)\\
&\tilde{A}_\theta(u,\textbf{r})=A_\theta(t,\textbf{r})\\
&\tilde{A}_\phi(u,\textbf{r})=A_\phi(t,\textbf{r})
\eea
\ee
One can see that only one of the components changes due to the coordinate transformation, that is, $\tilde{d}^{lm}(u,r)=d^{lm}(u+r^*,r)+b^{lm}(u+r^*,r)/f(r)$. Utilizing this we rewrite the components of the propagating EM wave as,
\be
\bea
&\tilde{A}_u(u,{\rb})=b^{lm}(u,r)Y_{lm}(\Omega)\\
&\tilde{A}_r(u,{\rb})=\tilde{d}^{lm}(u,r)Y_{lm}(\Omega)\\
&\tilde{A}_{s}(u,{\rb})=k_{lm}(u,r)\Psi^{lm}_{s}(\Omega)+a_{lm\mathrm{k}}(u,r)\Phi^{lm}_{s}(\Omega) .
\eea
\ee
From the equation of motion, $\nabla_\mu{F^{\mu\nu}}=0$, one gets the following coupled equation of the EM field propagating in Schwarzschild space time,
\be\label{coneq}
\bea
&\pr_r\tilde{\chi}^{lm}_1+\tilde{\chi}^{lm}_3=0\\
&\pr_u\tilde{\chi}^{lm}_1+f(r)\tilde{\chi}^{lm}_3-\tilde{\chi}^{lm}_4=0,
\eea
\ee
which can be further combined to obtain a decoupled equation of the following form 
\be
\pr_r(f(r)\pr_r\tilde{\chi}^{lm}_1)-2\pr_u\pr_r \tilde{\chi}^{lm}_1-\frac{l(l+1)}{r^2}\tilde{\chi}^{lm}_1=0 .
\ee
Also, the equation governing $\tilde{\chi}^{lm}_2$ turns out to be, 
\be\label{a1.eq}
\pr_r(f(r)\pr_r\tilde{\chi}^{lm}_2)-2\pr_u\pr_r \tilde{\chi}^{lm}_2-\frac{l(l+1)}{r^2}\tilde{\chi}^{lm}_2=0
\ee
where the expression for transformed invariant variables are
\be
\bea
&\tilde{\chi}^{lm}_1(u,r)=\frac{r^2}{l(l+1)}(\pr_u\tilde{d}^{lm}(u,r)-\pr_r b^{lm}(u,r))\\
&\tilde{\chi}^{lm}_2(u,r)=\tilde{a}^{lm}(u,r)=a^{lm}(u+r_*,r).
\eea
\ee
Other invariant variables can be computed from the constrained equation \eqref{coneq}.
\section{Derivation of eq.\eqref{numerator.def.abs} of the main text}\label{derv.numer.abs}
In this section we will explicitly derive the expression of eq.\eqref{numerator.def.abs} of the main text. Starting with the numerator of the definition \eqref{def1.abs}, we expand the the stress-energy tensor of the EM in terms of the gauge field components as follows,
\be
\bea
\pr_u\mathcal{F}&=\int{r^2d\Omega}{\mathcal{T}^r}_u\\
&=\int{d}\Omega{r^2}[{\mathcal{T}_{ru}}-{\mathcal{T}_{uu}}]\\
&=\int{d}\Omega{r^2}\Big[\Big\{-\frac{1}{4}g_{ru}F_{\mu\nu}F^{\mu\nu}+\frac{1}{2}g^{\nu\rho}(F_{r\nu}F_{u\rho}+F_{r\rho}F_{u\nu})\Big\}-\Big\{-\frac{1}{4}g_{uu}F_{\mu\nu}F^{\mu\nu}+g^{\nu\rho}F_{u\nu}F_{u\rho}\Big\}\Big]\\
&=\int{d}\Omega{r^2}\Big[-\frac{1}{4}(g_{ru}-g_{uu})F_{\mu\nu}F^{\mu\nu}+f(r)g^{\nu\rho}F_{r\nu}F_{u\rho}-g^{\nu\rho}F_{u\nu}F_{u\rho}\Big]\\
&=\int{d}\Omega{r^2}\Big[\Big(-g^{ur}(\pr_u\tilde{A}_r-\pr_r\tilde{A}_u)(\pr_u\tilde{A}^*_r-\pr_r\tilde{A}^*_u)\\
&+\frac{1}{2}g^{ss}(\pr_r\tilde{A}_s-\pr_s \tilde{A}_r)(\pr_u\tilde{A}^*_s-\pr_s\tilde{A}^*_u)+\frac{1}{2}g^{ss}(\pr_u\tilde{A}_s-\pr_s\tilde{A}_u)(\pr_r\tilde{A}^*_s-\pr_s\tilde{A}^*_r)\Big)\\
&-\Big(g^{rr}(\pr_u\tilde{A}_r-\pr_r\tilde{A}_u)(\pr_u\tilde{A}^*_r-\pr_r\tilde{A}^*_u)+g^{ss}(\pr_u\tilde{A}_s-\pr_s\tilde{A}_u)(\pr_u\tilde{A}^*_s-\pr_s\tilde{A}^*_u)\Big)\Big].\\
\eea
\ee
Where the ``tilde" index over the gauge field potential is due to the transformation from ($t,r$) to ($u,r$) as discussed in Appendix.\ref{eomucoord}. Now, we will substitute the mode decomposition of the gauge field as considered in the previous section.
\be
\bea
\pr_u\mathcal{F}&=\sum_{lm}\sum_{l'm'}\int{d}\Omega{r^2}\Big[\Big(-g^{ur}(\pr_u\tilde{d}^{lm}-\pr_r b^{lm})(\pr_u\tilde{d}^{l'm'^*}-\pr_r b^{l'm'^*})Y_{lm}(\Omega)Y^*_{l'm'}(\Omega)\\
&+\frac{1}{2}g^{ss}((\pr_r k^{lm}-\tilde{d}^{lm})\Psi^{lm}_{s}(\Omega)+\pr_r a^{lm}\Phi^{lm}_{s}(\Omega))((\pr_u k^{l'm'^*}-b^{l'm'})\Psi^{l'm'^*}_{s}(\Omega)+\pr_u a^{l'm'^*}\Phi^{l'm'^*}_{s}(\Omega))\\
&+\frac{1}{2}g^{ss}((\pr_u k^{lm}-b^{lm})\Psi^{lm}_{s}(\Omega)+\pr_u a^{lm}\Phi^{lm}_{s}(\Omega))((\pr_r k^{l'm'^*}-\tilde{d}^{l'm'})\Psi^{l'm'^*}_{s}(\Omega)+\pr_r a^{l'm'^*}\Phi^{l'm'^*}_{s}(\Omega))\Big)\\
&-\Big(g^{rr}(\pr_u\tilde{d}^{lm}-\pr_r b^{lm})(\pr_u\tilde{d}^{l'm'^*}-\pr_r b^{l'm'^*})Y_{lm}(\Omega)Y^*_{l'm'}(\Omega)\\
&+g^{ss}((\pr_u k^{lm}-b^{lm})\Psi^{lm}_{s}(\Omega)+\pr_u a^{lm}\Phi^{lm}_{s}(\Omega))((\pr_u k^{l'm'^*}-b^{l'm'^*})\Psi^{l'm'^*}_{s}(\Omega)+\pr_u a^{l'm'^*}\Phi^{l'm'}_{s}(\Omega))\Big)\Big]\\
\eea
\ee
One may notice that certain variables do not include a "tilde," as explained in the previous section. Utilizing the normalization of the vector spherical harmonics \cite{Barrera}, we integrate-out the angular part from the above equation. As a result, we obtain the following expression,   
\be
\bea
\pr_u\mathcal{F}&=\sum_{lm}{r^2}\Big[\Big\{-g^{ur}(\pr_u\tilde{d}^{lm}-\pr_r b^{lm})(\pr_u\tilde{d}^{lm^*}-\pr_r \tilde{b}^{lm^*})+\frac{l(l+1)}{2r^2}\Big((\pr_r k^{lm}-\tilde{d}^{lm})(\pr_u k^{lm^*}-b^{lm})\\
&+\pr_r a^{lm}\pr_u a^{lm^*}\Big)+\frac{l(l+1)}{2r^2}((\pr_u k^{lm}-b^{lm})(\pr_r k^{lm^*}-\tilde{d}^{lm})+\pr_u a^{lm}\pr_r a^{lm^*})\Big\}\\
&-\Big(g^{rr}(\pr_u\tilde{d}^{lm}-\pr_r b^{lm})(\pr_u\tilde{d}^{lm^*}-\pr_r b^{lm^*})+\frac{l(l+1)}{r^2}((\pr_u k^{lm}-b^{lm})(\pr_u k^{lm^*}-b^{lm^*})+\pr_u a^{lm}\pr_u a^{lm^*})\Big)\Big]\\
&=\sum_{lm}{r^2}\Big[\Big(\frac{l(l+1)}{2r^2}((\pr_r k^{lm}-\tilde{d}^{lm})(\pr_u k^{lm^*}-b^{lm})+\pr_r a^{lm}\pr_u a^{lm^*})\\
&+\frac{l(l+1)}{2r^2}((\pr_u k^{lm}-b^{lm})(\pr_r k^{lm^*}-\tilde{d}^{lm})+\pr_u a^{lm}\pr_r a^{lm^*})\Big)\\
&-\Big(\frac{l(l+1)}{r^2}((\pr_u k^{lm}-b^{lm})(\pr_u k^{lm^*}-b^{lm^*})+\pr_u a^{lm}\pr_u a^{lm^*})\Big)\Big]\\
\eea
\ee
In terms of the gauge invariant variables, defined in the ($u,r$) coordinate (see Appendix.\ref{eomucoord}), we finally arrive at,
\be
\bea
\pr_u\mathcal{F}&=\sum_{lm}\Big[\frac{l(l+1)}{2}\Big\{\Big(\tilde{\chi}^{lm}_3(u,r)\tilde{\chi}^{lm^*}_4(u,r)+\pr_r \tilde{\chi}^{lm}_2(u,r)\pr_u \tilde{\chi}^{lm^*}_2(u,r)\Big)\\
&+\Big(\tilde{\chi}^{lm}_4(u,r)\tilde{\chi}^{lm^*}_3(u,r)+\pr_u \tilde{\chi}^{lm}_2(u,r)\pr_r \tilde{\chi}^{lm^*}_2(u,r)\Big)\Big\}\\
&-l(l+1)\Big\{\tilde{\chi}^{lm}_4(u,r)\tilde{\chi}^{lm^*}_4(u,r)+\pr_u \tilde{\chi}^{lm}_2(u,r)\pr_u \tilde{\chi}^{lm^*}_2(u,r)\Big\}\Big]\\
\eea
\ee
This is exactly the expression that appears in eq.\eqref{numerator.def.abs} of the main text, with some added subscripts in the variables, due to the relevant spacetime region.

\end{document}